\documentclass{article}
\usepackage[T1]{fontenc}
\usepackage{lmodern}

\usepackage[italic]{mathastext}

\usepackage[english]{babel}
\usepackage[
    backend=biber,
    style=nature,
    doi=true,
    url=false
  ]{biblatex}
\AtEveryBibitem{\clearfield{month}}
\AtEveryBibitem{\clearfield{issn}}
\AtEveryBibitem{\clearfield{isbn}}

\usepackage[a4paper,
    top=2cm,bottom=2cm,left=3cm,right=3cm,marginparwidth=1.75cm
    ]{geometry}
\usepackage{graphicx}
\usepackage[colorlinks=true, allcolors=blue]{hyperref}

\usepackage[version=4]{mhchem}
\usepackage{amsmath}
\usepackage{amssymb}
\usepackage{textgreek}
\usepackage{siunitx}
\DeclareSIUnit{\molar}{M}
\DeclareSIUnit{\enzUnit}{U}
\DeclareSIUnit{\wtPercent}{wt\%}
\usepackage{nameref}
\usepackage{csquotes}
\usepackage{acronym}

\usepackage{titling}
\pretitle{\flushleft\huge} 
\posttitle{\par}
\preauthor{\flushleft\large}
\postauthor{\par}
\usepackage{authblk}

\title{Adaptive hydrogels with spatiotemporal stiffening using pH-modulating enzymes} 
\author[1,2]{Natascha Gray}
\author[1]{Zoe Grämiger}
\author[1,2]{André R. Studart}
\author[1,2 *]{Rafael Libanori}

\affil[1]{Complex Material, Department of Materials, ETH Zürich, 8093 Zürich, Switzerland}
\affil[2]{NCCR Bio-inspired Materials, ETH Zürich, 8093 Zürich, Switzerland \vspace{1\baselineskip}}
\affil[*]{Corresponding author}
\date{\vspace{-5ex}}

\acrodef{IQR}{interquartile range}
\acrodef{DN}{double network}
\acrodef{PAM}{polyacrylamide}
\acrodef{EDTA}{ethylenediaminetetraacetic acid}
\acrodef{CaEDTA}{calcium-ethylenediaminetetraacetic acid}
\acrodef{GOx}{glucose oxidase}
\acrodef{Fi}{ferricyanide}

\begin{document}
\maketitle

\section*{Abstract}
Biological systems achieve adaptive mechanical responses through reaction-diffusion processes that couple chemical wave propagation to structural transitions. Although synthetic hydrogels with enzymatic reactions offer a platform for replicating such autonomous behavior, the mechanistic principles governing chemomechanical transduction remain poorly understood. Here, we present a glucose oxidase-embedded polyacrylamide-alginate hydrogel with slower transduction kinetics that enable independent resolution of chemical waves and mechanical adaptation. Enzymatic pH waves propagate at 15–44 μm/min, triggering calcium-mediated alginate crosslinking through pH-responsive calcium-EDTA dissociation. Independent tracking of chemical and mechanical waves reveals that mechanical wavefronts (12 μm/min) lag behind chemical propagation, establishing transduction as the rate-limiting step in this chemomechanical coupling. Remarkably, the enzymatic system must continuously supply chemical energy to both propagate the chemical wave and drive ongoing mechanical transitions, imposing energetic costs on reaction-diffusion beyond kinetic constraints alone. Our adaptive system achieves up to 2.1-fold increase in stiffness and enables autonomous conversion of localized stimuli into system-wide mechanical responses. These mechanistic insights establish design principles for engineering adaptive materials with predictable spatiotemporal control in soft robotics and biomedical applications.
\acresetall


\section*{Introduction}
Adaptive hydrogels are an emerging class of materials that exhibit life-like functions typically associated with biological systems, such as sensing local environmental stimuli, amplifying and propagating chemical signals, and transducing these signals into coordinated spatiotemporal structural responses.\cite{walther2020ViewpointResponsive, chung2025SelfRegulatingHydrogel, merindol2017MaterialsLearning} Biological systems can achieve such spatiotemporal structural responses through reaction-diffusion processes, in which nonlinear chemical kinetics and diffusive transport generate propagating chemical waves that are transduced into structural responses. Examples include the rapid defense-stiffening of echinoderms\cite{wilkie2005MutableCollagenous} and the slow, controlled growth patterns of plants.\cite{martone2010MechanicsMuscle} Similarly, synthetic adaptive hydrogels can achieve autonomous behavior by harnessing chemical energy from these out-of-equilibrium reaction-diffusion processes to drive structural transitions in stimuli-responsive materials. By coupling reaction-diffusion signal output with the molecular responsiveness of hydrogel networks, spatiotemporal macroscopic transitions have been programmed into synthetic responsive materials through propagation of swelling-contracting waves,\cite{mao2020ContractionWaves} oscillating mechanical behavior,\cite{blanc2024CollectiveChemomechanical, yoshimura2020AutonomousOil, yoshida2010SelfOscillatingGels} and system-wide mechanical responses.\cite{paikar2022SpatiotemporalRegulation} Achieving predictable adaptive behavior in synthetic hydrogels requires elucidating the mechanistic coupling between chemical signals and structural transitions, particularly by identifying chemical reaction platforms that generate propagating chemical waves and activate spatiotemporal structural transitions in stimuli-responsive materials.\par
pH-modulating enzymatic reactions coupled with pH-responsive structural transitions provide a powerful integrated platform for implementing chemomechanical transduction mechanisms in hydrogel-based systems. Enzymes such as \ac{GOx}\cite{miguez2007FrontsPulses, vanag2006DesigningEnzymatic, fan2021PHFeedback} and urease\cite{duzs2024MechanoadaptiveMetagels, hu2010BaseCatalyzedFeedback, heuser2015BiocatalyticFeedback, jaggers2017TemporalSpatial} exhibit nonlinear kinetics that generate propagating pH waves capable of triggering structural transitions in pH-responsive hydrogels through supramolecular self-assembly of active species or dynamic crosslinking mechanisms. While urease-catalyzed reactions in low-viscosity or non-integrated liquid media achieve pH front speeds ranging from \qtyrange{10}{1000}{\um\per\minute},\cite{wrobel2012PHWaveFront,jee2016TemporalControl,heuser2017PhotonicDevices} hydrogel network reduces front velocities to $\sim$\qty{100}{\um\per\minute} in non-responsive hydrogels to \(<\)\qty{100}{\um\per\minute} in pH-responsive networks.\cite{duzs2024MechanoadaptiveMetagels} Mechanical response timescales for enzymatically-driven structural transitions also vary widely, from tens of minutes in enzyme‑mediated covalent crosslinking\cite{song2021RecentAdvancements, jonker2015FastActivatable} to several hours in fuel-driven transient hydrogels.\cite{sharma2023PHfeedbackSystems, boekhoven2015TransientAssembly} A recent study has shown that urease-driven pH wave fronts in a fully integrated hydrogel system promote fast spatiotemporal self-stiffening via pH-triggered supramolecular nanofiber assembly within pre-existing hydrogel networks.\cite{duzs2024MechanoadaptiveMetagels} Despite this progress, systematic investigation of the transduction pathways that convert chemical wave dynamics into programmed mechanical responses remains underexplored. To fully exploit the potential of these adaptive responses in real-world applications, it is necessary to elucidate the design principles governing these chemomechanical transduction mechanisms.\par
Chemomechanical transduction pathways can involve multiple physicochemical processes to convert chemical energy into mechanical modulation, including activation and diffusion of active species, self-assembly of small molecules, and dynamic crosslinking mechanisms. These transduction kinetics fundamentally determine the delay between chemical wave propagation and mechanical adaptation, which limits overall system responsiveness.\cite{geher-herczegh2021DelayedMechanical} State-of-the-art adaptive systems achieve self-stiffening through chemomechanical transduction mechanisms with varying response timescales. The urease-driven mechanoadaptive metagels reported by Dúzs and coworkers exemplify rapid transduction, in which measured pH wavefronts are transduced into spatiotemporally programmable mechanical responses following within 10-15 minutes.\cite{duzs2024MechanoadaptiveMetagels} While slower transduction kinetics facilitate systematic mechanistic investigation of chemomechanical transduction, they can impair material performance in applications that require rapid mechanical adaptation, such as soft robotic actuators\cite{lopez-diaz2024HydrogelsSoft} and biomedical valves for fluid-flow regulation.\cite{zheng2025MicropumpsMicrovalves} Conversely, slower transduction kinetics can offer distinct advantages in applications requiring more precise spatiotemporal control over mechanical adaptation response. For example, slow temporal mechanical modulation has been exploited to replicate the gradual increase in tumor stiffness over clinically relevant timescales.\cite{major2024ProgrammingTemporal} Therapeutic drug delivery platforms may require delayed structural transitions to trigger payload release at predetermined intervals, enabling staged drug delivery or sequential activation of chemical cascades.\cite{tan2017HeterogeneousMulticompartmental, pi2025EnzymeresponsiveBiomaterials} Therefore, elucidating transduction pathway kinetics and their coupling to chemical wave dynamics not only advances fundamental understanding of adaptive materials but also enables rational design of systems with application-matched response timescales.\par
Here, we present an adaptive \ac{DN} hydrogel system designed as a model platform to systematically investigate transduction pathway kinetics and their role in governing mechanical adaptation. The system achieves autonomous self-stiffening through direct enzymatic modulation of crosslinking density, in which \ac{GOx} generates acidic waves that propagate through the material, driving the progressive formation of Ca-alginate crosslinks within a \ac{PAM} matrix. The ionically crosslinked alginate network is rendered pH-responsive by adding \ac{EDTA}, which establishes a competitive equilibrium for \ce{Ca^2+} binding that shifts from CaEDTA to Ca-alginate following enzymatically driven alkaline-to-acidic pH transitions. Notably, our system's transduction kinetics, with mechanical waves following behind chemical waves, enable independent resolution of chemical signal propagation and mechanical response dynamics. By independently tracking chemical wave propagation using digital image processing and spatiotemporal mechanical stiffening kinetics through non-destructive indentation, we demonstrate that pH-triggered \ce{Ca^2+} release from EDTA complexes and subsequent diffusion-limited alginate crosslinking proceed on slower timescales than chemical wave propagation, establishing transduction pathway kinetics as the rate-limiting step in autonomous mechanical adaptation. Additionally, we also show that the enzymatic reaction must continuously supply chemical energy to both propagate the pH wave and sustain ongoing mechanical transduction, revealing that chemomechanical coupling can impose significant energetic costs on reaction-diffusion beyond kinetic constraints alone. These thermodynamic and kinetic mechanistic insights provide foundational design principles for engineering the next-generation autonomous adaptive materials with predictable, application-matched spatiotemporal mechanical adaptation.

\section*{Results and Discussion}
\begin{figure}[h!]
    \centering
    \includegraphics[width=\linewidth]{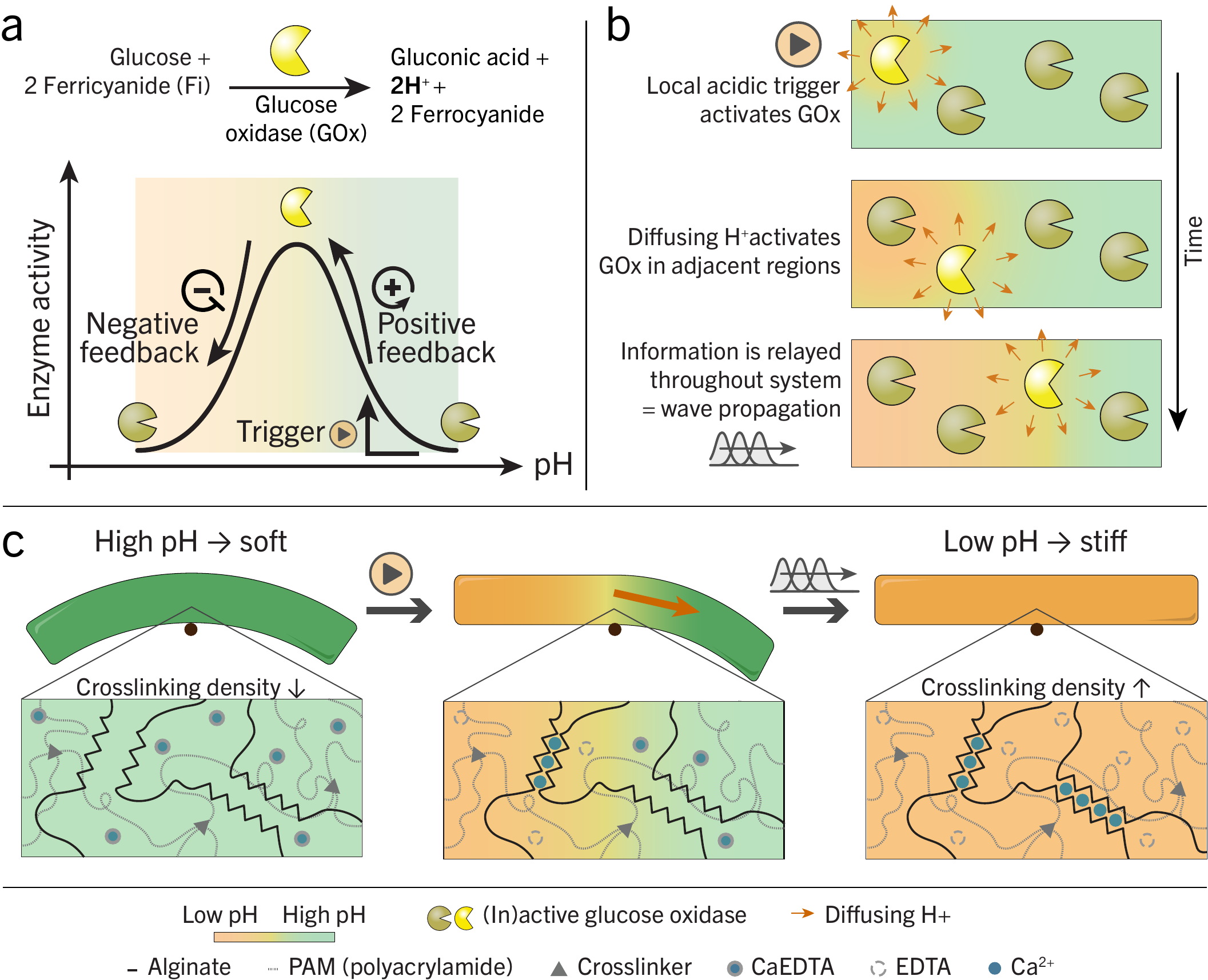}
    \caption{Design and mechanism of trigger-activated autonomous mechanical transition in adaptive \acf{DN} hydrogels. a) The \acf{GOx} enzymatic system and its pH-dependent activity profile. Top: \ac{GOx} catalyzes the oxidation of glucose to produce gluconic acid, decreasing the system's pH. Bottom: \ac{GOx}'s pH-dependent activity profile generates fast, autocatalytic positive feedback, followed by a negative feedback regulation, enabling alkaline-to-acidic pH transition. At high pH, \ac{GOx} exhibits minimal catalytic activity, maintaining the system in a quiescent (semi-dormant), excitable state that can be triggered by pH reduction prior to the spontaneous activation of \ac{GOx} autocatalysis. b) Spatiotemporal signal amplification and propagation from localized acidic triggering. Local acidic signal activates \ac{GOx}, initiating autocatalytic acid production that diffuses to sequentially activate adjacent \ac{GOx} molecules and generates self-sustained chemical waves. c) Conversion of chemical signals into mechanical transitions (chemomechanical transduction). The \acl{PAM}-alginate \ac{DN} hydrogel undergoes pH-dependent stiffening as \ac{EDTA} releases calcium ions (\ce{Ca^2+}) upon enzymatically-driven alkaline-to-acidic pH transition, enabling progressive increase in crosslinking density of alginate chains. The hydrogel transitions from a soft state (alkaline pH, \ce{Ca^2+} sequestered by EDTA) to a stiff state (acidic pH, Ca-alginate crosslinking).}
    \label{fig:idea}
\end{figure}

To achieve global mechanical stiffening from a local trigger, our system relies on the orchestrated integration of three key elements: nonlinear kinetics of pH-modulating enzymatic reactions, propagation of chemical signals through reaction-diffusion processes, and pH-responsive crosslinking within the hydrogel matrix (\textbf{\autoref{fig:idea}}). \Ac{GOx} is at the core of this integration, serving as both the engine powering mechanical transitions and the amplifier that propagates chemical signals throughout the adaptive system. \ac{GOx} catalyzes glucose oxidation to glucono-\textdelta-lactone, which spontaneously hydrolyzes to gluconic acid, thereby decreasing the pH of the reaction medium (\autoref{fig:idea}a and S1(i)-(ii)). This redox reaction reduces \ac{GOx} to its inactive form, which must be regenerated to its active, oxidized state by an electron-accepting mediator (Figure S1(iii)). While molecular oxygen \ce{O2} typically serves as the natural oxidizing agent for \ac{GOx} regeneration, it presents several experimental drawbacks such as poor water solubility, difficult control over concentration, and production of inhibitory \ce{H2O2}.\cite{wong2008GlucoseOxidase} Moreover, \ce{H2O2} formation in this redox reaction consumes two \ce{H+} per \ce{O2} molecule, reducing the amplification of the autocatalytic \ce{H+} production. These limitations motivate the use of alternative oxidizing agents for more reproducible reaction conditions.\cite{heller2008ElectrochemicalGlucose} Unlike \ce{O2}, \ac{Fi} enables precise control over initial conditions and does not consume additional \ce{H+} when regenerating \ac{GOx}, thereby enabling higher net \ce{H+} production rates and more rapid pH transitions. Additionally, the conversion of yellow \ac{Fi} to colorless ferrocyanide during \ac{GOx} oxidation enables direct optical monitoring of the enzymatic reaction state.\par
The pH-dependent activity profile of \ac{GOx} creates interconnected feedback regulations that enable autonomous pH modulation within the system (\autoref{fig:idea}a). At pH values above the optimum (ca. pH~6\cite{fan2021PHFeedback, weibel1971GlucoseOxidase, keilin1948PropertiesGlucose}), \ac{GOx} produces acidic species that increase its activity, creating an autocatalytic positive feedback auto-regulation that amplifies acid production. Conversely, at pH values below the optimum, continued acid production generates a negative feedback auto-regulation in which \ac{GOx} activity progressively decreases as the out-of-equilibrium reaction approaches steady-state (at ca. pH 2-3\cite{fan2021PHFeedback, weibel1971GlucoseOxidase, keilin1948PropertiesGlucose}). This non-linear pH-dependent response allows the system to be kinetically trapped into a quiescent excitable state, wherein the autocatalytic positive feedback remains in a semi-dormant state. Under these conditions, external acidic triggers can activate this excitable state by decreasing the pH below a critical threshold, initiating a spontaneous alkaline-to-acidic transition throughout the system.\par 
In addition to exploiting \ac{GOx}'s pH-dependent activity for temporal alkaline-to-acidic transitions, we harness reaction-diffusion dynamics to implement spatial responses in the system. The coupling of \ac{GOx}'s nonlinear reaction and diffusion kinetics generates self-sustained chemical waves in the reaction medium, thereby enabling the implementation of spatially-modulated pH transitions in our adaptive system (\autoref{fig:idea}b). Localized addition of an acidic trigger activates \ac{GOx} in a limited volume, generating \ce{H+} that diffuse outward and sequentially activate \ac{GOx} in adjacent regions. This generates a self-sustained chemical wave front that propagates through the entire system, converting a local chemical signal into a global chemical response. Unlike passive diffusion fronts that decay with distance, these wave fronts can propagate at much higher speeds and maintain constant amplitude over centimeter-scale distances.\cite{merindol2017MaterialsLearning, heuser2017PhotonicDevices} Because their propagation speed is determined by the combination of diffusive transport and non-linear enzymatic kinetics, spatiotemporal transitions can be modulated by varying \ac{GOx} concentration, substrate and mediator availability, and the system's buffer capacity.\cite{miguez2007FrontsPulses, kovacs2010FrontPropagation} The spatiotemporal pH signal, propagated by the enzymatically-driven chemical wave, enables global mechanical transitions when coupled with responsive hydrogels that exhibit pH-dependent crosslinking density (\autoref{fig:idea}c). We employ a \ac{DN} hydrogel comprised of permanently crosslinked \ac{PAM} and reversibly crosslinked calcium-alginate (\ac{PAM}-CaEDTA-alginate).\cite{zhang2022GradientAdhesion} The \ac{PAM} network provides structural integrity, while the Ca-alginate network introduces crosslinking dynamics. Addition of the chelating agent \ac{EDTA} renders this network pH-responsive by establishing a pH-dependent competitive equilibrium for calcium binding. As pH decreases, progressive protonation of EDTA's carboxylic acid groups (p$K$\textsubscript{a1} = 2.0; p$K$\textsubscript{a2} = 2.7; p$K$\textsubscript{a3} = 6.2; p$K$\textsubscript{a4} = 10.3),\cite{raaflaub1956ApplicationsMetal} reduces EDTA's affinity for \ce{Ca^2+}, causing \ce{Ca^2+} ions to preferentially bind to the carboxylate groups of guluronic acid residues in alginate chains,\cite{onsoyen1997Alginates} thereby increasing crosslinking density. Embedding the \ac{GOx} enzymatic reaction in the pH-responsive \ac{DN} hydrogel yields a metastable, semi-dormant, adaptive system capable of sensing, amplifying, and propagating localized chemical signals to induce autonomous global mechanical transitions.\par

\begin{figure}[h!]
    \centering
    \includegraphics[width=\linewidth]{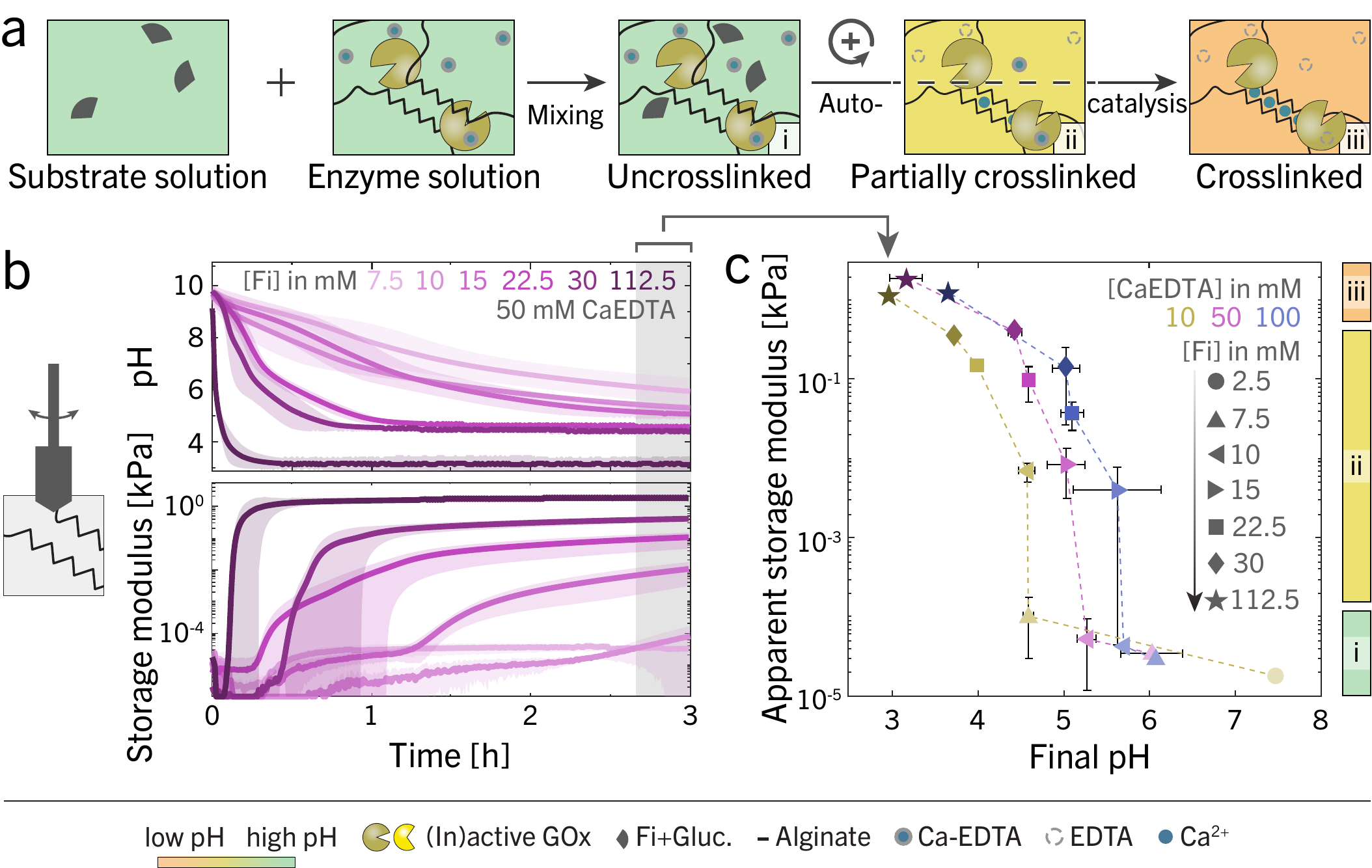}
    \caption{Spontaneous autonomous mechanical transition in pH-responsive alginate induced by \acf{GOx} autocatalysis. a) Time-sweep rheology with in situ pH monitoring measurements of autonomous temporal mechanical transitions in \acf{CaEDTA}-alginate system. Stock solutions containing \acf{Fi} and CaEDTA-alginate are mixed immediately before measurement, enabling monitoring of storage modulus evolution during the \ac{GOx}-driven pH transition. b) Averaged pH transition and time-sweep curves for \ac{Fi} concentrations ranging from \qtyrange{7.5}{112.5}{\milli\molar} at constant CaEDTA and glucose (\qty{50}{\milli\molar} and \qty{100}{\milli\molar}, respectively). c) Apparent storage modulus as a function of the final pH. Symbols and colors represent different \ac{Fi} and CaEDTA concentrations, respectively; color scale on the right side indicates alginate crosslinking state from (a). Shaded areas in (b) represent standard deviation. Error bars in (c) show standard deviation of the values acquired in the final \qty{20}{\minute} of measurement.}
    \label{fig:mechRheo}
\end{figure}

The alkaline-to-acidic pH transition induced autocatalytically by the embedded \ac{GOx} can be harnessed to activate the crosslinking mechanism of the pH-responsive CaEDTA-alginate network (\textbf{Figure \ref{fig:mechRheo}}). Upon spontaneous activation of the \ac{GOx} enzymatic reaction, \ce{Ca^2+} ions are continuously released from CaEDTA complexes and bind to guluronic acid residues in adjacent alginate chains, progressively increasing the material's storage modulus and inducing a sol-gel transition. To monitor the evolution of this process during the autonomous pH transition, we employ oscillatory rheology with in situ pH measurement (\autoref{fig:mechRheo}, \hyperref[method:rheopH]{Experimental Section}). Time-sweep experiments were performed on samples prepared by mixing two stock solutions immediately before testing (\autoref{fig:mechRheo}a). One solution contains substrate ([glucose] = \qty{100}{\milli\molar}) and mediator (\qty{7.5}{\milli\molar} $\leq$ [\ac{Fi}] $\leq$ \qty{112.5}{\milli\molar}) while the other contains enzyme ([\ac{GOx}] = \qty{100}{\enzUnit\per\gram}), uncrosslinked polymer (\qty{2}{\wtPercent} alginate) and the pH-responsive crosslinking supplier ([CaEDTA] = \qtylist{10;50;100}{\milli\molar}). The glucose concentration is maintained well above the Michaelis-Menten constant for \ac{GOx}-catalyzed glucose oxidation ($K$\textsubscript{m} = \qtyrange{2.4}{18.0}{\milli\molar}),\cite{zhou2005GlucoseBiosensor, shan2008AmperometricGlucose, arslan2014AmperometricBiosensor} ensuring that the enzymatic rate depends solely on the \ac{GOx} and \ac{Fi} concentrations. Additionally, the initial pH was set to \num{10.0 \pm 0.1} to minimize \ac{GOx} activity and favor CaEDTA complex formation while minimizing the risk for irreversible \ac{GOx} denaturation and \ce{Fe(OH)3} precipitation at more alkaline pH values. These conditions ensured that the CaEDTA-alginate system transitions from uncrosslinked to crosslinked states during the measurement.\par 
Time-sweep measurements were limited to \qty{3}{\hour}, providing sufficient time for most samples to approach equilibrium storage modulus ($G'_{eq}$) and pH values. We defined the apparent storage modulus ($G'_{app}$) and final pH as the averaged data from the final \qty{20}{\minute} of measurement (the shaded region depicted in \autoref{fig:mechRheo}b and summarized in \autoref{fig:mechRheo}c). It should be noted that sample compositions with slow reaction kinetics may not reach equilibrium within the 3-hour rheological measurement window, potentially underestimating their actual $G'_{eq}$ values. Under conditions of excess substrate and constant \ac{GOx} and CaEDTA concentrations, the final pH and $G'_{app}$ can be tuned by varying the initial \ac{Fi} concentration (\autoref{fig:mechRheo}b). At constant CaEDTA (\qty{50}{\milli\molar}), increasing \ac{Fi} from \qty{7.5}{\milli\molar} (lightest pink curve in \autoref{fig:mechRheo}b) to \qty{112.5}{\milli\molar} (darkest pink curve) shifts the final pH from \num{6.0 \pm 0.4} to \num{3.2 \pm 0.2}. Because lower pH values favor the crosslinked state of the CaEDTA-alginate network, the $G'_{app}$ increases by five orders of magnitude, from \qty{3.4 \pm 0.2e-5}{\kilo\pascal} to \qty{1.9 \pm 0.1}{\kilo\pascal}. Additionally, higher \ac{Fi} concentrations increase both the system's capacity to restore \ac{GOx}'s active state and its regeneration turnover rate, significantly reducing the time required to reach final pH levels from \qty{3}{\hour} to \qty{0.5}{\hour} across the \ac{Fi} concentration range explored in this study.\par
The pH-dependent evolution of $G'_{app}$ reveals three distinct crosslinking states for CaEDTA-alginate systems. The lower plateau at more alkaline pH values corresponds to an uncrosslinked network state (green region indicated on the lateral color bar in \autoref{fig:mechRheo}c), whereas the upper plateau values indicate a crosslinked state (orange region) at pH values which approach the inherent lower limit determined by the negative feedback regulation exerted by the \ac{GOx} pH-dependent activity.\cite{fan2021PHFeedback, weibel1971GlucoseOxidase, keilin1948PropertiesGlucose} CaEDTA-alginate systems with $G'_{app}$ values between these plateau states represent a partially crosslinked transition state (yellow region). Since the fraction of \ce{Ca^2+} released from CaEDTA at any given pH remains constant regardless of total CaEDTA concentration, systems with higher CaEDTA concentrations provide a larger absolute amount of free \ce{Ca^2+}. This higher \ce{Ca^2+} availability increases the maximum $G'_{app}$ achievable at a given pH and shifts the crosslinking onset to more alkaline pH values (\autoref{fig:mechRheo}c). Based on this pH-dependent analysis, we estimated that the onset of crosslinking shifts from approximately pH 4.6 to 5.3 to 5.6 as CaEDTA concentration increases from \qty{10}{\milli\molar} (yellow dataset in \autoref{fig:mechRheo}c) to \qty{50}{\milli\molar} (pink dataset) to \qty{100}{\milli\molar} (blue dataset). The pH interval ranging from approximately 3.2–6.0 aligns well with the reported pH interval where \ce{Ca^2+} preferential binding shifts from EDTA to alginate.\cite{bassett2016CompetitiveLigand}\par
While this pH-dependent crosslinking mechanism suggests that increasing CaEDTA concentrations should enhance the mechanical transition through increased \ce{Ca^2+} availability, the $G'_{app}$ exhibits divergent trends with CaEDTA concentration depending on initial [\ac{Fi}] (\autoref{fig:mechRheo}c). At \qty{22.5}{\milli\molar} \ac{Fi}, the $G'_{app}$ decreases from \qty{1.5 \pm 0.1d-1}{\kilo\pascal} to \qty{9.8 \pm 4.7d-2}{\kilo\pascal} to \qty{3.7 \pm 1.5e-2}{\kilo\pascal} with increasing CaEDTA concentration from \qty{10}{\milli\molar} to \qty{50}{\milli\molar} to \qty{100}{\milli\molar} (Figure \ref{fig:SIrheoCE} and square symbols in \autoref{fig:mechRheo}c). In contrast, we observed a non-monotonic increase with a maximum in $G'_{app}$ for higher \ac{Fi} concentrations (\qty{30.0}{\milli\molar}, diamond symbols; \qty{112.5}{\milli\molar}, star symbols). This complex behavior arises from the antagonistic dual role of CaEDTA. While serving as the \ce{Ca^2+} source for alginate crosslinking, EDTA acts as a strong \ce{H+} acceptor that buffers pH changes.\cite{raaflaub1956ApplicationsMetal} Because the conditional stability constant ($K'_f$) of CaEDTA is directly proportional to the fraction of deprotonated EDTA, lower pH values reduce its \ce{Ca^2+} binding affinity, thereby releasing more free \ce{Ca^2+} for crosslinking. Consequently, higher CaEDTA concentrations increase buffering capacity, which limits pH reduction and maintains more \ce{Ca^2+} in the CaEDTA complex, ultimately reducing crosslinking density. These results illustrate the intricate interplay between enzyme kinetics, ion release, and acid-base equilibria in CaEDTA-alginate network formation.\par

\begin{figure}[h!]
    \centering
    \includegraphics[width=\linewidth]{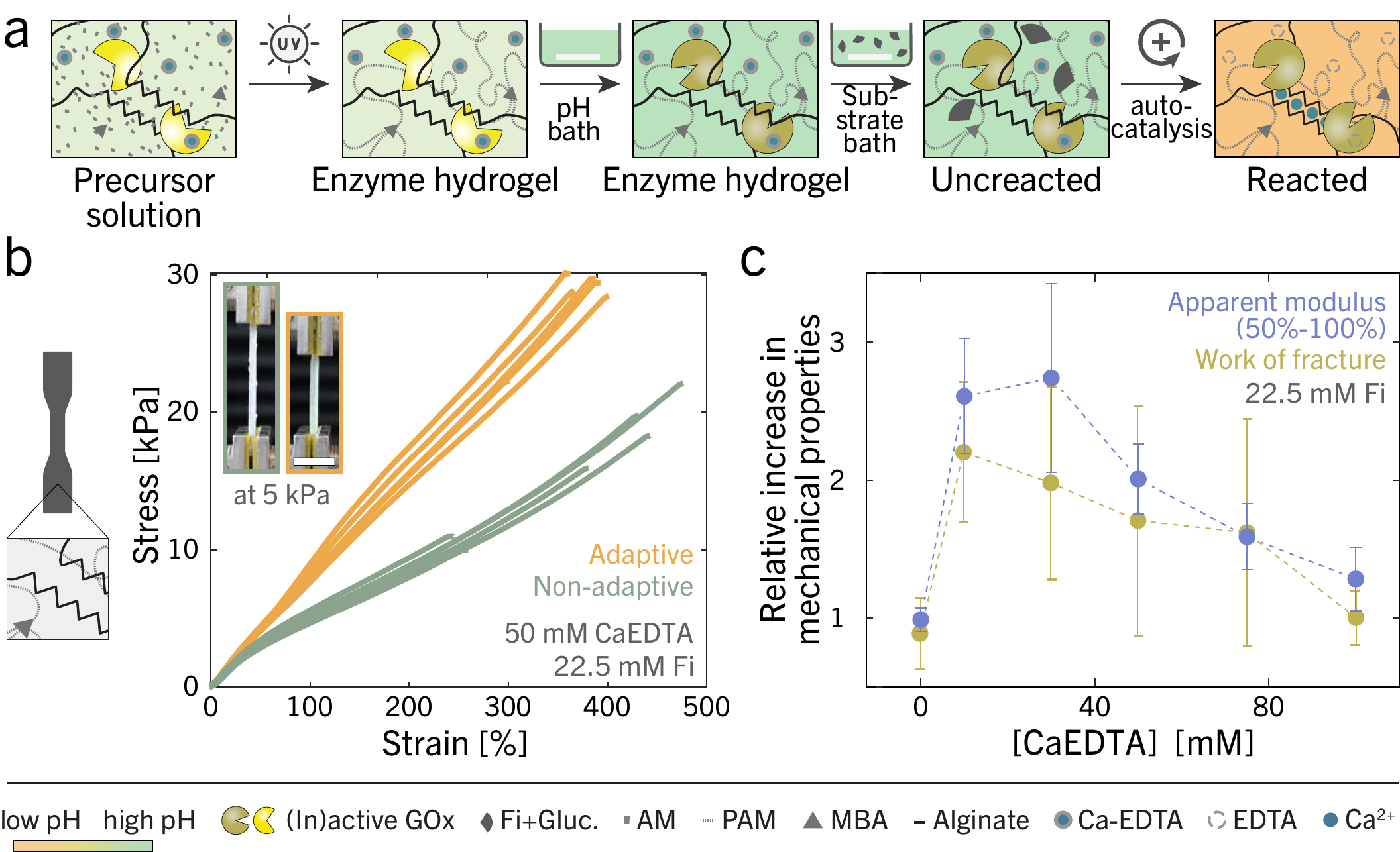}
    \caption{Spontaneous autonomous mechanical transition in pH-responsive \acf{DN} alginate-based hydrogels induced by \acf{GOx} autocatalysis. a) Experimental protocol for quasi-static tensile testing of \acf{PAM}-\acf{CaEDTA}-alginate \ac{DN} hydrogels. The \ac{PAM} network is crosslinked with bisacrylamide (MBA) in the presence of uncrosslinked alginate and CaEDTA, followed by substrate and mediator incorporation via by bath immersion. Tensile tests are performed after \ac{GOx}-driven pH transition (adaptive) or without substrate/mediator (non-adaptive). b) Stress-strain curves for adaptive and non-adaptive samples (\qty{50}{\milli\molar} CaEDTA, \qty{22.5}{\milli\molar} \acl{Fi}, \qty{100}{\milli\molar} glucose, $n = 6$). The scale bar shows \qty{1}{\cm}. c) Relative increase in apparent modulus and work of fracture of adaptive and non-adaptive samples. Apparent modulus is determined from a linear fit of the \qtyrange{50}{100}{\percent} strain interval; work of fracture is calculated from the area under the stress-strain curves. Error bars in (c) represent the standard deviation calculated via statistical error propagation. Difference between mechanical properties of the adaptive and non-adaptive systems is statistically significant (as defined in the Experimental Section for \(n=6\) or 7), with the exception of [CaEDTA] = \qty{0}{\milli\molar} and the work of fracture at  [CaEDTA] = \qty{100}{\milli\molar}.}
    \label{fig:mechTens}
\end{figure}

The implementation of an enzymatically driven crosslinking mechanism that dynamically forms a reinforcing network in the presence of a permanently crosslinked network enables the design of adaptive \ac{DN} hydrogels that exhibit self-stiffening behavior (\textbf{\autoref{fig:mechTens}}). To demonstrate this, we adapted existing protocols for \ac{DN} hydrogels synthesis to fabricate \ac{PAM} networks (\qty{12.5}{\wtPercent}) via free-radical photopolymerization in the presence of alginate (\qty{2}{\wtPercent}), \ac{GOx} (\qty{100}{\enzUnit\per\gram}), and CaEDTA (\qtyrange{0}{100}{\milli\molar}) (\autoref{fig:mechTens}a).\cite{hirsch20213DPrinting, sun2012HighlyStretchable} Because the photopolymerization process decreases pH and decomposes \ac{Fi}, and including glucose during polymerization could initiate premature enzymatic activity, we incorporated the substrate and mediator by diffusion after hydrogel synthesis. First, we immersed the as-synthesized hydrogels for \qty{10}{\minute} in a bath containing a solution at pH \num{10.0 \pm 0.1} with CaEDTA concentration equivalent to the sample composition. Subsequently, substrate and mediator were incorporated by immersing the samples for \qty{10}{\minute} in a second CaEDTA solution at a pH of \num{10.0 \pm 0.1} containing glucose and \ac{Fi} at concentrations of \qty{100}{\milli\molar} and \qty{22.5}{\milli\molar}, respectively. After incorporating the substrate and mediator, the hydrogels were immersed in mineral oil overnight to allow spontaneous autocatalytic pH decrease driven by \ac{GOx}, thereby inducing Ca-alginate crosslink formation and network stiffening. \par
The mechanical reinforcement resulting from spontaneous \ac{GOx}-driven crosslinking was quantified via quasi-static uniaxial tensile testing, comparing stress-strain responses of adaptive hydrogels (orange curve in) to non-adaptive controls (green curve). Samples were rendered non-adaptive by excluding glucose from their formulations and they thus remained at pH 10 until tested. The stress-strain curves reveal that non-adaptive hydrogels exhibit more pronounced nonlinear elastic response between \qtyrange[range-phrase=~--~]{0}{50}{\percent} applied strain, whereas adaptive hydrogels display a significant enhancement in mechanical response due to enzymatic activity. Remarkably, the strain at rupture is not affected by the spontaneous autonomous mechanical transition. To exclude non-linear behavior from the quantitative analysis, we extracted the apparent modulus by calculating the slope of the stress-strain curve between \qtyrange[range-phrase=~--~]{50}{100}{\percent} of the applied strain, and the work of fracture by integrating the area under the stress-strain curves. Adaptive hydrogels containing CaEDTA and \ac{Fi} at concentrations \qty{50}{\milli\molar} and \qty{22.5}{\milli\molar}, respectively, exhibited a \num{2.0 \pm 0.3}-fold increase in the apparent modulus and a \num{1.7 \pm 0.8}-fold increase in the work of fracture compared to the reference hydrogels. The self-stiffening effect in adaptive hydrogels is also evident when comparing the relative deformation experienced by the reacted and reference samples at the same applied stress (inset images in \autoref{fig:mechTens}b).\par
The antagonistic role of CaEDTA in the pH-dependent crosslinking mechanism of the CaEDTA-alginate network is also reflected in the mechanical response of \ac{DN} hydrogels in quasi-static tensile mode. The variation of relative increase in apparent modulus and work of fracture with increasing CaEDTA concentration clearly demonstrates the interplay between the pH-dependent competitive mechanism for \ce{Ca^2+} binding and the buffering effect of EDTA (\autoref{fig:mechTens}c). At CaEDTA concentrations below \qty{30}{\milli\molar}, the low buffer capacity does not significantly affect the acid-base and binding equilibria, and CaEDTA serves primarily as a source of \ce{Ca^2+} ions for the crosslinking mechanism. As a result, the relative increase in apparent modulus is directly proportional to the amount of CaEDTA, reaching a maximum value of \num{2.7 \pm 0.7} at \qty{30}{\milli\molar}. Conversely, at concentrations above \qty{30}{\milli\molar}, the high buffering capacity of CaEDTA dominates the acid-base equilibria and dampens the pH-modulating driving force provided by the enzymatic reaction. In this regime, increasing CaEDTA concentration results in a continuous decrease in apparent modulus, reaching values only \num{1.3 \pm 0.2}-fold higher than non-adaptive hydrogels. Notably, the apparent modulus and work of fracture of non-adaptive hydrogels do not exhibit such a strong dependence on CaEDTA concentration and the apparent modulus of non-adaptive hydrogels at constant free \ce{Ca^2+} concentration is only weakly pH-dependent (Figure \ref{fig:SItens} and \ref{fig:SItenspH}). These findings confirm that the observed trends arise from pH-dependent release of \ce{Ca^2+} rather than direct effects of CaEDTA on the PAM network or a pH dependence of the Ca-alginate crosslinking. Importantly, the mechanical response of the adaptive \ac{DN} hydrogel measured in quasi-static loading mode is consistent with that of the CaEDTA-alginate single network in oscillatory rheology (dynamic loading mode), suggesting that the \ac{PAM} network does not significantly impact the pH-responsive crosslinking process of the CaEDTA-alginate network. Surprisingly, the adaptive \ac{DN} hydrogels reported here show a more modest increase in mechanical performance than previous studies of similar \ac{DN} hydrogels, which reported up to a 6-fold and 3.5-fold increase in elastic modulus and strain at rupture, respectively.\cite{sun2012HighlyStretchable, gorke2024UnravellingParameter} While significant, the mechanical reinforcement achieved in our adaptive \ac{DN} hydrogels is constrained by limited pH reduction, which affects network development through two mechanisms. First, the competitive binding equilibrium between EDTA and alginate restricts \ce{Ca^2+} availability for intra-network ionic crosslinking , confirmed by quasi-static tensile testing of non-adaptive \ac{DN} hydrogels without EDTA which showed significantly higher apparent moduli (Figure \ref{fig:SItenspH}). Second, low protonation degree of alginate carboxylate groups weakens both intra-network hydrogen bonds between alginate chains and inter-network hydrogen bonds with PAM. More acidic pH conditions are expected to enhance mechanical performance by simultaneously increasing \ce{Ca^2+} release for ionic crosslinking and promoting carboxylate protonation, thereby strengthening hydrogen bonding interactions. The use of weaker chelating agents, such as ethylenediaminediacetic acid (EDDA), could shift the \ce{Ca^2+} binding equilibrium toward alginate, while incorporating additional autocatalytic \ce{H+} sources could enable more acidic pH conditions and thereby increase the availability of free \ce{Ca^2+} for alginate crosslinking. Overall, this dual-network strategy demonstrates that temporal pH regulation via enzymatic catalysis enables autonomous mechanical transitions, producing adaptive \ac{DN} hydrogels that self-stiffen through enzymatically-controlled alginate crosslinking.\par

\begin{figure}[h!]
    \centering
    \includegraphics[width=\linewidth]{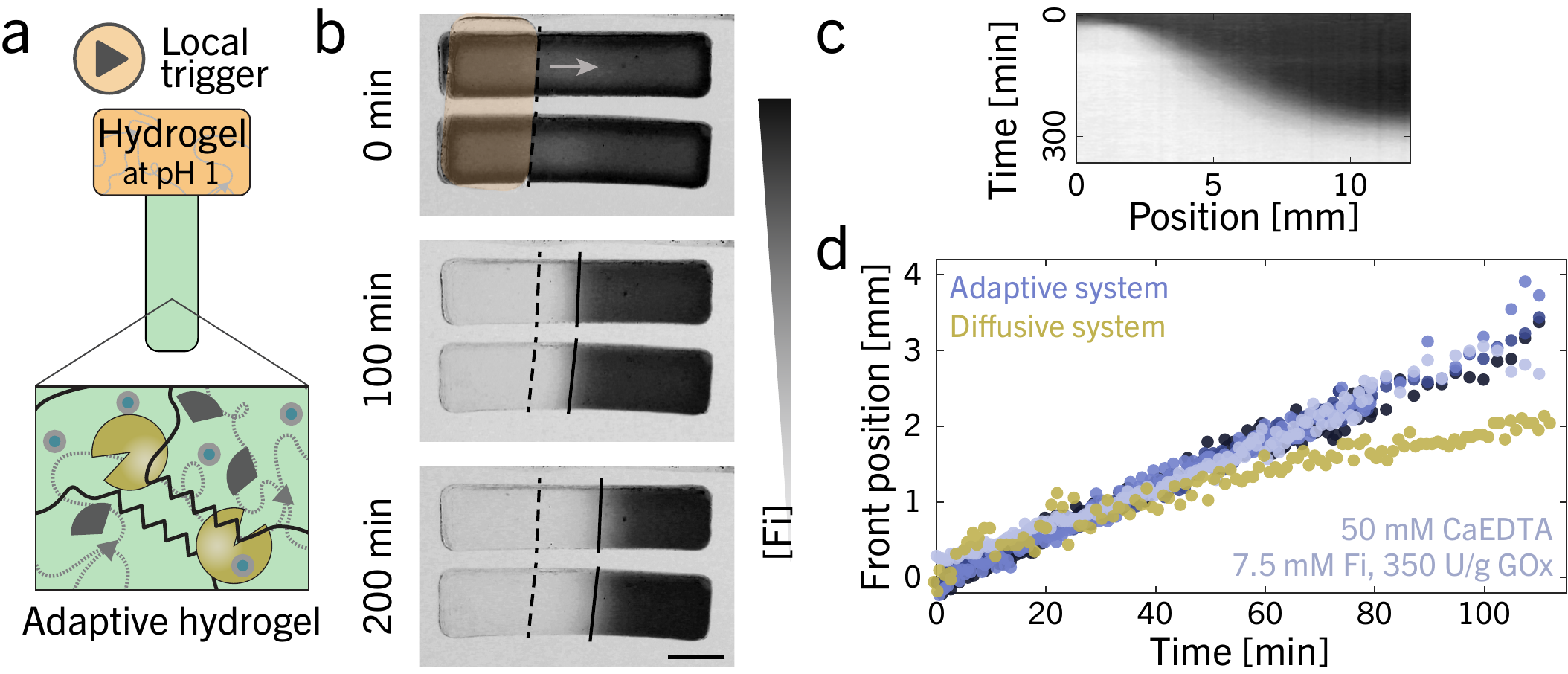}
    \caption{Propagation of self-sustained chemical waves in \acf{DN} hydrogels. a) Experimental protocol for initiating chemical wave propagation in adaptive \ac{DN} hydrogels. A local chemical signal is delivered by contacting the \ac{DN} hydrogel with a \acl{PAM} hydrogel pre-equilibrated at pH 1 for at least \qty{24}{\hour}. b) Representative blue channel images showing duplicate samples at three time points. The trigger hydrogel is false-colored orange ($t = \qty{0}{\minute}$) and defines the triggered region (dashed line), with the propagating chemical wave front marked by a solid line. Sample compositions: \acl{GOx} (GOx, \qty{250}{\enzUnit\per\gram}), \acl{Fi} (Fi, \qty{22.5}{\milli\molar}), \acl{CaEDTA} (CaEDTA, \qty{50}{\milli\molar}), \acl{PAM} (\qty{12.5}{\wtPercent}) and alginate (\qty{2}{\wtPercent}). Scale bar: \qty{5}{\mm}. c) Representative kymograph showing the spatiotemporal progression of the self-sustained chemical wave in the adaptive \ac{DN} hydrogel. d) Temporal evolution of wave front positions in four adaptive \ac{DN} hydrogel samples from the same batch, presented in shades of blue. For comparison, the front position of a purely diffusive system (yellow) is shown, determined by Alizarin Red S color change upon 1-minute contact with a hydrogel at pH 1. Sample compositions: adaptive system contained \ac{GOx} (\qty{350}{\enzUnit\per\gram}), \ac{Fi} (\qty{7.5}{\milli\molar}), glucose (\qty{100}{\milli\molar}), and CaEDTA (\qty{50}{\milli\molar}); diffusive system contained CaEDTA (\qty{50}{\milli\molar}), both contained \acl{PAM} (\qty{12.5}{\wtPercent}) and alginate (\qty{2}{\wtPercent}). The trigger hydrogel was removed at $t = \qty{0}{\minute}$ for all experiments.}
    \label{fig:waveProp1}
\end{figure}

\begin{figure}[h!]
    \centering
    \includegraphics[width=\linewidth]{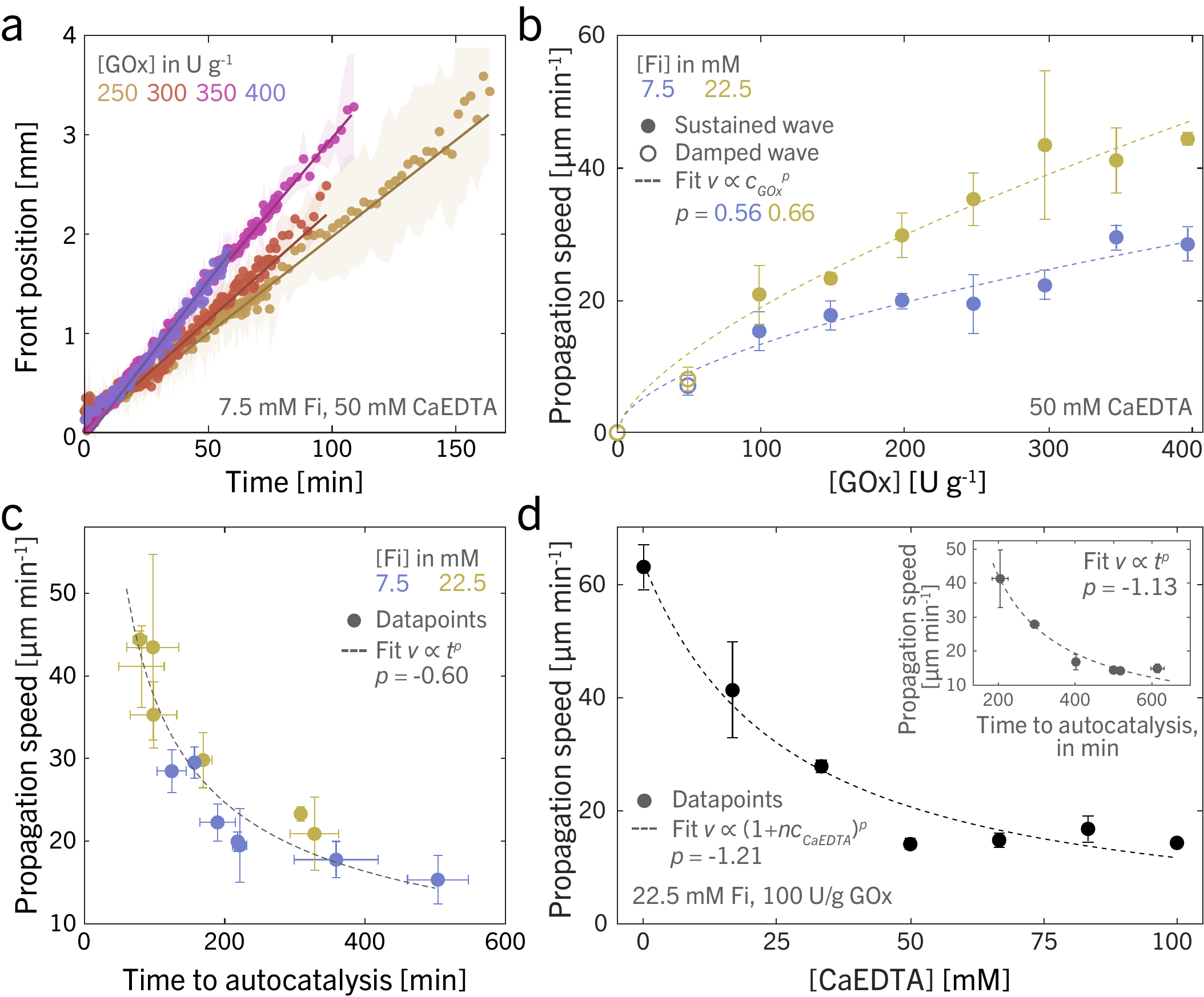}
    \caption{Chemical wave propagation kinetics in adaptive \acf{DN} hydrogels. a) Temporal evolution of wavefront positions in adaptive hydrogels containing varying \acf{GOx} concentrations (\qty{250}, \qty{300}, \qty{350}, and \qty{400}{\enzUnit\per\gram}) at fixed \acl{Fi} (Fi, \qty{7.5}{\milli\molar}) and \acl{CaEDTA} (CaEDTA, \qty{50}{\milli\molar}). Data points represent mean values whereas shaded regions indicate the standard deviation ($n = 4$), difference between distance of lowest (\qty{250}{\enzUnit\per\gram}) and highest (\qty{400}{\enzUnit\per\gram}) GOx concentration mentioned in the text is statistically significant for $n=4$. b)~Wave propagation speed versus \ac{GOx} concentration for two \ac{Fi} concentrations (\qty{7.5}{\milli\molar} and \qty{22.5}{\milli\molar}), at constant CaEDTA (\qty{50}{\milli\molar}). Solid and open circles indicate self-sustained and damped wave propagation, respectively. c) Correlation between propagation speed and time to spontaneous autocatalysis (time-to-autocatalysis) for the dataset in (b). d) Propagation speed versus CaEDTA concentration at fixed \ac{Fi} (\qty{22.5}{\milli\molar}) and \ac{GOx} (\qty{100}{\enzUnit\per\gram}). Inset shows propagation speed versus time-to-autocatalysis. Dashed lines in (b–d) represent power law fits ($v \propto x^p$) with exponents indicated in the legends. All adaptive DN hydrogels contained a fixed glucose (\qty{100}{\milli\molar}), \acl{PAM} (\qty{12.5}{\wtPercent}) and alginate (\qty{2}{\wtPercent}) concentrations.}
    \label{fig:waveProp2}
\end{figure}

Beyond inducing spontaneous temporal pH transitions, the chemical energy released by out-of-equilibrium enzymatic reactions can be harnessed to propagate self-sustained chemical waves in \acf{DN} hydrogels (\ac{PAM}-CaEDTA-alginate) (\textbf{Figure \ref{fig:waveProp1}} and \textbf{\ref{fig:waveProp2}}). To demonstrate this potential, we prepared adaptive \ac{PAM}-CaEDTA-alginate strips measuring \qtyproduct[product-units=power]{2 x 5 x 20}{\mm} using the experimental protocol depicted in \autoref{fig:mechTens}a. Freshly synthesized samples were immersed in mineral oil to prevent water evaporation and minimize \ce{O2} diffusion, as \ce{O2} can compete with \ac{Fi} as \ac{GOx} mediator in certain reaction conditions.\cite{wang2008ElectrochemicalGlucose} Rather than relying on spontaneous \ac{GOx} autocatalysis to drive the pH transition, a \ac{PAM} hydrogel equilibrated at pH~1 was brought into contact with one end of the adaptive hydrogel for durations ranging from \qtyrange{1}{5}{\minute} (\autoref{fig:waveProp1}a and \autoref{fig:waveProp1}b at $t = \qty{0}{\minute}$). This time-limited physical contact transfers \ce{H+} by diffusion from the acidic trigger hydrogel to the alkaline adaptive hydrogel, lowering the local pH and activating the semi-dormant \ac{GOx}. Following an induction period, ranging from \qty{0.5}{\minute} to \qty{20}{\minute} (\autoref{fig:waveProp1}c), inversely correlated with contact time the system reaches a new dynamic steady state, primarily governed by \ac{GOx} enzymatic activity.\par 
We exploited the conversion of yellow \ac{Fi} into colorless ferrocyanide caused by GOx regeneration to optically monitor the spatiotemporal enzymatic reaction coordinate (Figure \ref{fig:SIimgA}, \hyperref[method:optTrack]{Experimental Section}). Time-lapsed RGB images captured by a conventional reflex camera were analyzed using a MATLAB script that extracts the blue channel intensity and converts it into grayscale. The reaction front position was approximated as the inflection point of the grayscale intensity profile along the sample's longitudinal axis (Figure \ref{fig:SIimgA}). Our results reveal the emergence of a reaction front whose position scales linearly with time until spontaneous activation of \ac{GOx} autocatalysis ahead of the propagating front $t_{autocatalysis} = \qty{270}{\minute}$ in \autoref{fig:waveProp1}c), which induces a global color transition of the sample. The time-to-autocatalysis is defined as the inflection point of the temporal grayscale intensity variation of samples that underwent spontaneous GOx autocatalysis (Figure \ref{fig:SIimgA}). In contrast to purely diffusive fronts, \ac{PAM}-CaEDTA-alginate hydrogels containing \ac{GOx} (\qty{350}{\enzUnit\per\gram}), \ac{Fi} (\qty{7.5}{\milli\molar}), and CaEDTA (\qty{50}{\milli\molar}) exhibited chemical fronts propagating at a constant speed of \qty{30 \pm 2}{\um\per\minute} between $t = \qty{0}{\minute}$ and $t = \qty{120}{\minute}$ (\autoref{fig:waveProp1}d) throughout the PAM-CaEDTA-alginate hydrogel network. Despite its multi-component composition, the adaptive \ac{DN} system exhibited high reproducibility, with a maximum coefficient of variability of $0.11$ across both intra- and inter-batch measurements (Figure \ref{fig:waveProp1}d and \ref{fig:SIinterbatch}). These self-sustained chemical fronts arise from the reaction-diffusion mechanism governed by the non-linear reaction and diffusion kinetics of the enzymatic system.\cite{miguez2007FrontsPulses} Self-sustained propagation is further corroborated by experiments where propagation speed remains largely unaffected when trigger time varies from \qtyrange{0.16}{15}{\minute} (Figure \ref{fig:SItrigger}a). In contrast, the front position in purely diffusive systems strongly depends on trigger time (Figure \ref{fig:SItrigger}b). Quantification of apparent \ce{H+} diffusion coefficient from purely diffusive propagation fronts reveals that increasing trigger time from \qtyrange{0.16}{15}{\minute} also increases the apparent \ce{H+} diffusion coefficient from \qty{1.3\pm 0.4d-6}{\cm\squared\per\second} to \qty{8.8\pm 1.9d-6}{\cm\squared\per\second} (Figure \ref{fig:SItrigger}b). Note that the absolute numbers depend on the choice of indicator, these numbers are therefore only accurate in a relative sense. Self-sustained wave propagation eventually terminates (\autoref{fig:waveProp1}c) because the semi-dormant GOx undergoes homogeneous autocatalytic activation ahead of the wave front, thereby limiting the maximum propagation distance. Extending wave propagation distance requires establishing a deeper dormant enzyme state through more extreme alkaline initial pH conditions that suppress spontaneous \ac{GOx} autocatalysis. However, we limited the initial pH value to \num{10.0\pm 0.1} to improve system reproducibility and stability, as highly alkaline conditions favors \ac{Fi} instability and irreversible GOx deactivation. Alternatively, strategies that do not rely solely on kinetic balancing of initial conditions, such as incorporating an extrinsic enzymatic regulatory mechanism, could establish a truly excitable dormant state through active regulation rather than extreme initial conditions. Such approaches could enable higher initial substrate concentrations, potentially allowing the enzymatic system to overcome buffering effects, suppress spontaneous nucleation, achieve faster propagation speeds, and extend propagation range while improving overall system stability.\par

The propagation speed of chemical waves across the adaptive \ac{DN} hydrogel is modulated by the enzymatic \ce{H+} production rate and the system's buffering capacity (\autoref{fig:waveProp2}). The \ce{H+} production rate is controlled by GOx and Fi concentrations at constant glucose concentration (\qty{100}{\milli\molar}), while the system's buffering capacity is dominated by the CaEDTA concentration. Propagation speeds were determined from linear regression of wavefront positions versus time for at least four independent intra-batch measurements per composition (solid dark lines in \autoref{fig:waveProp2}a). The results reveal that increasing \ac{GOx} concentration causes propagation speed to rise asymptotically (\autoref{fig:waveProp2}b), reaching a maximum value of \qty{44 \pm 1}{\um\per\minute} within the design space explored in this study. The observed propagation speeds of \qty{15 \pm 3}{\um\per\minute} to \qty{44 \pm 1}{\um\per\minute} were substantially lower than those reported for other hydrogel systems, ranging from \qtyrange{100}{2000}{\um\per\minute}.\cite{duzs2024MechanoadaptiveMetagels, miguez2007FrontsPulses} However, reference measurements in single-network hydrogels composed of \ac{PAM} alone yielded propagation speeds on the order of \qty{1200}{\um\per\minute}, which aligns well with previously reported values for similar networks containing \ac{GOx} enzymatic reactions mediated by \ac{Fi} (Figure \ref{fig:SIpam}a)\cite{miguez2007FrontsPulses} and validates our experimental approach. The markedly reduced propagation speeds in our \ac{DN} hydrogel system likely arise from the combination of the strong buffering capacity of EDTA and the presence of alginate chains within the \ac{PAM} network, which provide additional buffering and affect the local water structure, significantly altering \ce{H+} diffusion mechanisms within the hydrogel by suppressing the fast hopping found in bulk water (Figure \ref{fig:SIpam}b).\cite{choi2005ThermodynamicsProton, feng2011ProtonSolvation, mabuchi2019EffectsWater} Additionally, the increase in network tortuosity may slow the diffusion of counterions and other small molecules.\par
To obtain quantitive insights on the effects of the the enzymatic \ce{H+} production rate and system's buffering capacity on the chemical wave propagation speed, we applied a theoretical model describing the propagation kinetics of chemical waves generated by autocatalytic \ce{H+} production in monoprotic buffered systems.\cite{kovacs2010FrontPropagation} This model (\textbf{\autoref{eq:model}}) predicts that propagation speed $v$ scales with autocatalytic rate ($k$), initial substrate concentration ($s$), and buffer concentration ($c_b$):
\begin{equation}
    v\propto \sqrt{ksD_h}/(1+c_bK_b)
    \label{eq:model}
\end{equation}
where $D_h$ is the effective \ce{H+} diffusion coefficient, and $K_b$ is the buffer equilibrium constant. Assuming that under constant concentrations of glucose and \ac{Fi}, $k$ is proportional to \ac{GOx} concentration, and $c_b$ positively correlates to CaEDTA concentration, the model predicts a square-root dependence (exponent $p = 0.5$) on \ac{GOx} concentration and inverse scaling ($p = -1$) with the CaEDTA concentration. We tested these predictions by fitting experimental data with power law functions: $v=a\cdot c_{GOx}{}^p$ for \ac{GOx}-dependent propagation speed and $v=a\cdot (1+n\cdot c_{CaEDTA})^p$ for CaEDTA-dependent propagation speed, where $a$, $n$ and $p$ are fitting parameters. The experimental scaling exponents obtained from our datasets are 0.56 and 0.66 for \ac{Fi} concentrations of \qty{7.5} and \qty{22.5}{\milli\molar}, respectively (\autoref{fig:waveProp2}b), in reasonable agreement with that predicted by the model ($p = 0.5$). This power law relationship confirms that propagation speed is positively correlated with local \ce{H+} production rate, which is constrained by GOx turnover under mediator-limited conditions. The scaling exponent increases from 0.56 to 0.66 with a 3-fold increase in \ac{Fi} concentration from \qty{7.5}{\milli\molar} to \qty{22.5}{\milli\molar}, demonstrating that higher mediator concentrations enhance the sensitivity of propagation speed to \ac{GOx} concentration. At the lower \ac{Fi} concentration of \qty{7.5}{\milli\molar}, the maximum propagation speed is limited to only \qty{28 \pm 3}{\um\per\minute}, representing a \qty{-35}{\percent} reduction compared to the higher Fi condition (\qty{22.5}{\milli\molar}) (\autoref{fig:waveProp2}b).\par
The eventual spontaneous activation of \ac{GOx} autocatalysis imposes a temporal constraint on propagation dynamics, limiting the maximum distance chemical waves can travel within the adaptive \ac{DN} hydrogel (\autoref{fig:waveProp2}a,c). Because both the propagation speed and time required for spontaneous autocatalysis activation are regulated by local \ce{H+} production rate, increasing \ac{GOx} concentration simultaneously accelerates wave propagation and shortens autocatalysis activation time (Figure \ref{fig:waveProp2}b,c and \ref{fig:SIautocat2}). The higher absolute scaling exponents obtained from power law fitting of time-to-autocatalysis ($p = -0.96$ and $-1.08$ in Figure \autoref{fig:SIautocat2}b) compared to propagation speed ($p = 0.56$ and $0.66$ in \autoref{fig:waveProp2}b) suggests that \ac{GOx} concentration exerts a stronger impact on autocatalysis activation than on wave propagation speed. This asymmetric scaling relationship establishes an inverse correlation between propagation speed and time-to-autocatalysis, with the difference in scaling exponents ($p_{time-to-autocatalysis}-p_{propagation\,speed}\cong -0.5$) indicating that autocatalysis activation is approximately twice as sensitive to \ac{GOx} concentration than propagation speed (\autoref{fig:waveProp2}c). Because the maximum propagation distance is proportional to time-to-autocatalysis, faster waves are more spatially confined due to rapid autocatalytic activation compared to slower waves, which can propagate longer distances. For instance, wavefronts propagating at \qty{28 \pm 3}{\um\per\minute} cover distances of only \qty{1.6 \pm 0.1}{\mm} (purple dataset in \autoref{fig:waveProp2}a), whereas slower waves propagating at \qty{19 \pm 4}{\um\per\minute} extend the maximum propagation distance by \qty{100}{\percent} (\qty{3.2 \pm 0.8}{\mm}, yellow dataset). Thus, maximizing wave propagation distance requires balancing two opposing constraints. At low \ac{GOx} concentrations, self-sustained propagation fails because diffusive and buffering losses exceed the local \ce{H+} production rate. According to our experiments, the critical \ac{GOx} concentration required to generate a self-sustained wave propagation lies between \qty{50}-\qty{100}{\enzUnit\per\gram} (\autoref{fig:waveProp2}b). In contrast, at high \ac{GOx} concentrations, rapid spontaneous autocatalytic activation limits propagation time, confining waves to shorter distances despite faster speeds. Therefore, intermediate \ac{GOx} concentrations establish a kinetic regime in which only propagation speed or propagation distance can be maximized individually, yielding slow and long-lasting chemical waves at \ac{GOx} concentrations near the critical lower limit, or fast and short-lived chemical waves at high \ac{GOx} concentrations. \par
While \ac{GOx} and \ac{Fi} modulate local \ce{H+} production rate to promote propagation of self-sustained chemical waves, CaEDTA exerts an antagonistic effect through buffering, also consistent with theoretical model predictions.\cite{kovacs2010FrontPropagation} Compared to CaEDTA-free systems where propagation speed reaches \qty{63\pm 4}{\um\per\minute}, the introduction of \qtyrange{16}{100}{\milli\molar} CaEDTA reduces the propagation speed to only \qty{14\pm 1}{\um\per\minute} (\autoref{fig:waveProp2}d). Power-law fitting of propagation speed versus CaEDTA concentration ($v=a\cdot (1+n\cdot c_{CaEDTA})^p$) yields a scaling exponent of $p = -1.21$, which is comparable to that predicted by the model and reflects the strong buffering capacity that EDTA exerts on the adaptive \ac{DN} hydrogel. The reasonable agreement between model predictions and experimental observations demonstrates that the theoretical framework captures the essential physicochemical mechanisms governing wave propagation, despite several simplifying assumptions. Most notably, the model assumes pH-independent autocatalytic kinetics, suggesting that \ac{GOx}'s pH-dependent activity does not significantly affect wave propagation dynamics in our system.  This approximation proves valid because autocatalytic front speed is determined by reaction conditions at the leading edge rather than the trailing edge.\cite{vansaarloos2003FrontPropagation} While pH gradients behind the wavefront are steep and affect \ac{GOx} activity significantly, \ce{H+} concentration gradients at the leading edge remain shallow, resulting in minimal variation in \ac{GOx} activity across this region. Thus, the effective \ac{GOx} autocatalytic rate governing wave propagation remains relatively constant across the pH conditions present at the leading edge, validating the pH-independent approximation. Additionally, while CaEDTA exhibits polyprotic acid-base equilibria, the model's monoprotic buffer approximation appears adequate for predicting the overall buffering effect on wave speed. These results validate the use of simplified reaction-diffusion frameworks for understanding enzymatic wave propagation in adaptive \ac{DN} hydrogels. The integration of tunable spatiotemporal wave propagation dynamics (Figure \ref{fig:waveProp1} and \ref{fig:waveProp2}) with programmable mechanical self-stiffening (Figure \ref{fig:mechRheo} and \ref{fig:mechTens}) demonstrates the potential of using these adaptive \ac{DN} hydrogels as a model platform for designing adaptive materials that sense localized chemical cues, amplify and propagate these signals through responsive networks, and execute spatially programmed soft-to-stiff mechanical transitions.

\begin{figure}[h!]
    \centering
    \includegraphics[width=\linewidth]{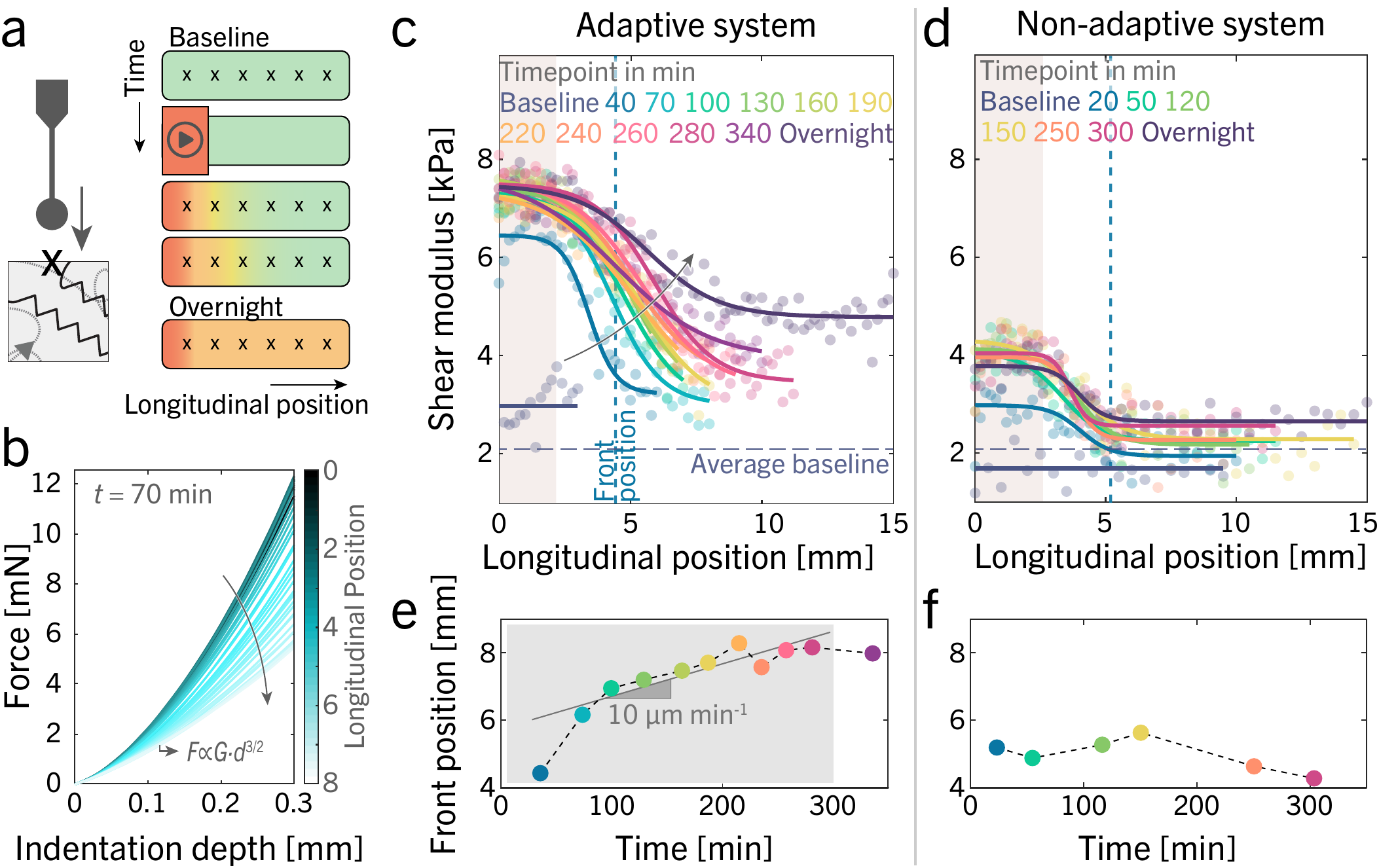}
    \caption{Spatiotemporal mechanical wave propagation in adaptive \acf{DN} hydrogels. a) Experimental protocol for spatiotemporally resolved macroindentation measurements. Shear modulus values, $G$, are determined by repeated indentation along the longitudinal axis at regular intervals following baseline characterization. b) Representative spatially resolved force-indentation depth curves obtained at $t = \qty{70}{\minute}$. $G$ was calculated by Hertzian contact mechanics. Evolution of spatially resolved $G$ over time for adaptive (c) and non-adaptive (d) \ac{DN} hydrogels. The mechanical wavefront position is tracked by extracting the inflection points of sigmoidal curves fitted for each temporal profile (continuous lines). Dashed horizontal lines indicate the average $G$ baseline measured before mechanical wave activation. Temporal evolution of the mechanical wavefront position in adaptive (e) and non-adaptive (f) \ac{DN} hydrogels. The mechanical wave propagation speed is determined from the pre-autocatalysis phase (gray region in e), where self-sustained chemical waves were observed to propagate with constant speed. The adaptive \ac{DN} hydrogel system (c,e) contains \acl{GOx} (\qty{100}{\enzUnit\per\gram}), \acl{Fi} (\qty{22.5}{\milli\molar}), glucose (\qty{100}{\milli\molar}), and \acl{CaEDTA} (\qty{30}{\milli\molar}), whereas the non-adaptive \ac{DN} hydrogel system (d,f) lacks all enzymatic reaction components, both systems contain \acl{PAM} (\qty{12.5}{\wtPercent}) and alginate (\qty{2}{\wtPercent}).}
    \label{fig:mechTrack}
\end{figure}

The mechanical wave propagates along the hydrogel longitudinal axis following chemical wave activation, producing distinct spatiotemporal patterns in self-stiffening response (\textbf{\autoref{fig:mechTrack}}). We quantified these spatiotemporal mechanical transitions through non-destructive indentation measurements along the longitudinal axis of adaptive and non-adaptive \ac{DN} hydrogel samples (\autoref{fig:mechTrack}a). Adaptive hydrogels contained \ac{GOx}, \ac{Fi}, and glucose to enable enzymatically driven pH transitions, while non-adaptive reference systems consisted of the \ac{DN} hydrogel matrix alone. Following the preparation and triggering protocols described previously  (\autoref{fig:mechTens}a and \autoref{fig:waveProp1}a), samples were subjected to consecutive spatially resolved indentation measurements along their longitudinal axis using a \qty{1.5}{\mm} radius probe at time intervals ranging from \qty{20}{\minute} to overnight (\autoref{fig:mechTrack}b). Initial mechanical states (baseline) were quantified by performing spatially resolved indentations on each sample before chemical wave activation. Each spatiotemporal measurement yielded a force-indentation depth curve from which the shear modulus ($G$) was calculated using Hertzian contact mechanics and corrected for sample thickness effects (\hyperref[method:mechTrack]{Experimental Section}).\cite{shull2002ContactMechanics}\par
The spatial $G$ profiles acquired at each time interval after triggering the chemical wave display characteristic sigmoidal shapes that evolve over time (\autoref{fig:mechTrack}c). These $G$ profiles exhibit a maximum plateau in the triggered region due to \ce{H+} diffusion from the acidic trigger hydrogel (shaded region in \autoref{fig:mechTrack}c). In the proximal (triggered) region, the $G$ achieves \qty{7.0}{\kilo\pascal} (\ac{IQR} \qtyrange[range-phrase=~--~]{5.8}{7.6}{\kilo\pascal}), representing a 1.3-fold and 3.3-fold increase over the non-adaptive system (\qty{5.3}{\kilo\pascal}, \ac{IQR} \qtyrange[range-phrase=~--~]{4.50}{5.6}{\kilo\pascal}) and the average baseline (\qty{2.1}{\kilo\pascal}, \ac{IQR} \qtyrange[range-phrase=~--~]{1.8}{2.3}{\kilo\pascal}), respectively (Figure \ref{fig:SImTG}, incl. statistical significance). Moving toward the distal region, the $G$ profiles gradually transition to a minimum plateau value of \qty{4.5}{\kilo\pascal} (\ac{IQR} \qtyrange[range-phrase=~--~]{3.2}{5.0}{\kilo\pascal}), representing a 2.1-fold increase compared to the average baseline. This mechanical reinforcement level matches the $G$ of \qty{4.5}{\kilo\pascal} (\ac{IQR} \qtyrange[range-phrase=~--~]{3.8}{5.6}{\kilo\pascal}) presented by equilibrated non-triggered adaptive samples after spontaneous \ac{GOx} autocatalysis (Figure \ref{fig:SImTG} and \ref{fig:SImTref}), suggesting that enzymatic activity alone determine the mechanical reinforcement in the distal region.\par
Surprisingly, the distal region in triggered non-adaptive hydrogels exhibits an unexpected self-stiffening with $G$ being \qty{57}{\percent} above baseline values, yet remains \qty{27}{\percent} lower than the distal self-stiffening induced by spontaneous \ac{GOx} autocatalysis (triggered non-adaptive references, Figure \ref{fig:mechTrack}d, \ref{fig:SImTG} and \ref{fig:SImTref}). Although these samples were exposed to the acidic triggering hydrogel, this unexpected stiffening in distal regions cannot be attributed to \ce{H+} diffusion from the proximal region, as \ce{H+} transport in \ac{DN} hydrogel networks is too slow to induce mechanical changes at these timescales. Such slow \ce{H+} transport kinetics is confirmed by preliminary diffusion experiments in single- and \ac{DN} hydrogels. Adaptive \ac{DN} hydrogels exhibit persistent pH gradients even $\sim$1~day after being triggered by the acidic hydrogel (Figure \ref{fig:SIpHLongEDX}a). In addition to graded pH profiles, adaptive and non-adaptive samples also exhibited detectable persistent mechanical gradients for at least 6~days after removal of the acidic triggering hydrogel (Figure \ref{fig:SIpHLongEDX}b). This persistent gradient likely stems from the spatially varying \ce{H+} concentration rather than from \ce{Ca^2+} accumulation due to CaEDTA mobility following \ce{Ca^2+} immobilization. Energy-dispersive X-ray spectroscopy measurements reveal no statistically significant Ca concentration difference between proximal and distal ends, while pH indicator measurements confirm a persistent pH gradient across the sample (Figure \ref{fig:SIpHLongEDX}a,c). This observation highlights both the strong pH dependence in the mechanical properties as well as the slow kinetics of local crosslinking formation. Since slow \ce{H+} diffusion, as discussed previously, cannot account for the unexpected self-stiffening in the reference samples, we hypothesize that such an increase in $G$ in the distal region of reference samples may be caused by improved sample-substrate bedding caused by repeated indentations and/or by strain-induced network rearrangements and water migration, which are facilitated by the dynamic nature of CaEDTA-alginate crosslinks.\cite{wang2023UnderstandingGelPowers, zhang2007ProposedStrainhardening} This hypothesis is corroborated by the observation that non-adaptive \ac{DN} hydrogels which are not placed in contact with the acidic triggering hydrogel also present a significant self-stiffening, achieving a 1.75-fold increase in $G$ over a period of $\sim \qty{4}{\hour}$ being subjected to repeated indentation measurements (Figure \ref{fig:SImTref}b).\par
In addition to hosting the enzymatic reaction, the \ac{DN} hydrogel functions as a chemomechanical transducer, converting the self-sustained chemical waves into spatiotemporal soft-to-stiff mechanical transitions. To quantify this chemomechanical coupling, we tracked the mechanical wavefront position over time by determining the leading front of the transition at the onset of the distal plateau of the $G$ profiles during the quiescent period between trigger removal and spontaneous \ac{GOx} autocatalytic activation (time-to-autocatalysis in Figure \autoref{fig:waveProp2}c, \qtyrange{0}{300}{\minute} for all, without baseline). For this, we utilized a robust fitting algorithm designed to mitigate the influence of significant experimental outliers. Such deviations are inherent to the complexity of these experimental systems (Figure \ref{fig:SImTSpeedsFull}a-b). For instance, the data point at \(t=\qty{40}{\minute}\) in Figure \ref{fig:mechTrack}e suggests an artificially accelerated initial phase; however, this represents an isolated artifact not reproduced in other experiments (Figure \ref{fig:SImTSpeedsFull}a). In adaptive systems, the mechanical wavefront advances significantly with a median propagation speed of \qty{12}{\um\per\minute} (\ac{IQR} \qtyrange{10}{23}{\um\per\minute}), whereas it remains stationary in non-adaptive reference systems (\qty{2}{\um\per\minute} IQR \qtyrange{-1}{4}{\um\per\minute}, Figure \ref{fig:mechTrack}e-f and \ref{fig:SImTSpeedsFull}c). Remarkably, the mechanical wavefront propagates approximately \qty{40}{\percent} times slower than the speed of the chemical wave determined by optical tracking at the same experimental conditions (Figure \ref{fig:SImTSpeedsFull} and \ref{fig:waveProp2}), suggesting that chemomechanical transduction leads to a kinetic lag in the propagation of mechanical waves. Such kinetic lag is likely caused by the temporal offset between the chemical transition and the delayed structural response that is often observed in other autonomous structural transitions.\cite{heuser2017PhotonicDevices} Specifically, the delayed mechanical response of these adaptive \ac{DN} hydrogels is caused by the multiple-step crosslinking mechanism inherent to the dynamic CaEDTA-alginate network. Following enzymatic-induced acidification, \ce{Ca^2+} must be released from CaEDTA, diffuse to alginate chains to promote ionic crosslinking, and accumulate to reach levels that can be detectable by indentation.\par
The demonstrated integration of out-of-equilibrium enzymatic reactions with pH-responsive crosslinking dynamics establishes a versatile platform for programmable chemomechanical transduction in soft materials. This integrated platform enables autonomous sensing of localized chemical cues, signal amplification through autocatalytic feedback regulation, spatial communication via self-sustained chemical wave propagation, and spatiotemporally programmable mechanical transitions. Future work could extend this adaptive platform by incorporating enzymatic cascade reactions to create more complex signal-processing networks, implementing signal transduction strategies to detect other types of external stimuli, integrating extrinsic regulatory mechanisms to stabilize excitable states and extend wave-propagation distances, or coupling chemical and mechanical transitions to other functional outputs such as controlled release of small molecules, autonomous self-healing, or adaptive mechanical actuation. The reversible nature of CaEDTA-alginate crosslinks further enables the integration of reversible autonomous mechanical transitions, allowing adaptive systems to be reset and re-triggered multiple times provided chemical fuel is continuously supplied and waste removed. Overall, this work demonstrates how reaction-diffusion dynamics and reversible crosslinking chemistry can be orchestrated to create hydrogel-based materials with life-like capabilities, providing insights into relevant design principles for advancing truly intelligent synthetic systems capable of autonomous and programmable structural transitions.\par

\section*{Conclusion}
\acresetall
We have developed model adaptive hydrogels capable of autonomous spatiotemporal mechanical transitions by integrating a pH-modulating enzyme, \ac{GOx}, into a pH-responsive \acl{PAM}-alginate \ac{DN} hydrogel. This autonomous material system offers unique mechanistic insights into chemomechanical transduction mechanisms by harnessing chemical energy from out-of-equilibrium enzymatic reactions to drive programmable spatiotemporal stiffening through localized chemical signal sensing, amplification and propagation. The enzymatic reaction-diffusion process generates self-sustained chemical waves with propagation speeds tunable between \qtyrange[range-phrase=~--~]{15}{44}{\um\per\minute} through systematic control of enzymatic kinetics and system's buffering capacity, following predictable power-law scaling relationships with \ac{GOx} and \acl{Fi} concentrations, and inverse scaling with CaEDTA concentration. Crucially, the adaptive system achieves chemomechanical transduction by converting propagating chemical signals into traveling mechanical stiffening fronts. This transduction occurs through a multi-step cascade, consisting of enzymatically-induced pH decrease, pH-triggered \ce{Ca^2+} release from CaEDTA complexes, \ce{Ca^2+} diffusion to alginate chains, and a progressive increase in crosslinking degree. The mechanical front propagates at approximately \qty{12}{\um\per\minute}, which is \qty{40}{\percent} slower than the chemical wave, revealing the slower kinetic timescales of \ce{Ca^2+}-mediated crosslinking that govern the transduction process. This chemomechanical coupling drives autonomous material stiffening, increasing the shear modulus by up to 2.1-fold and establishing mechanical gradients that persist for days due to slowed \ce{H+} diffusion within the strongly buffered \ac{DN} hydrogel. While a remaining gradient in mechanical properties may limit the applicability of such a system, it offers enticing possibilities in some areas, such as tissue engineering or drug delivery. By integrating chemical sensing, autocatalytic signal amplification, wave-mediated signal transmission, and chemomechanical transduction within a single programmable material system, this work provides key mechanistic insights into chemomechanical coupling required for the development of next-generation autonomous, adaptive materials with predictable, application-matched spatiotemporal mechanical adaptation.

\textbf{Corrigendum} \par
The authors wish to clarify that the pH-dependent activity of glucose oxidase critically depends upon the specific mediator employed. While the citations in the text pertain to enzyme activities observed with oxygen, which exhibit a bell-shaped pH-dependence with a maximum at pH 6, the use of ferricyanide in this study results in a monotonic increase in activity as pH decreases to 3.\cite{wilson1992GlucoseOxidase} Preliminary experimental data based on liquid phase experiments confirms this monotonic increase in activity down to pH 4, but shows an attenuated and incomplete ferricyanide turnover at pH 3. Furthermore, embedding glucose oxidase in a hydrogel may further alter the observed trend, as the surrounding chemical environment can strongly affect pH-dependent activity. Consequently, the negative feedback inherent to the enzyme is expected to be negligible for this system, but stem from mediator depletion instead. 

\bigskip
\textbf{Acknowledgments} \par
The authors would like to express their heartfelt gratitude to Dr. Hanspeter Schmid for his open and willing assistance in automating the linear fitting to measure the propagation speed and the statistical analysis. Our sincere gratitude to Dr. Robert Style for opening the world of indentation measurements and analysis to us. We also thank Dr. Kirill Feldmann for introducing the pH rheology setup and measurement, Nicolas Wyler for his support in the optical tracking experiments, Christian Furrer for producing the in-situ rheology setup and the cooling plate for the UV curing chamber, Dr. Elena Tervoort for the SEM and EDX analysis, and finally Prof. Nico Bruns, Dr. Luciano Boesel, and Sam Russell for the fruitful discussions. This work was financially supported by the Swiss National Science Foundation through the NCCR Bio-inspired Materials (51NF40-205603).

\bigskip
\textbf{Author Contributions} \par
R.L. and N.G. conceptualized the research project. R.L. and N.G. designed the experiments. N.G. and Z.G. carried out the experiments: N.G. performed rheology, tensile testing, and indentation, with Z.G. providing support in developing the indentation protocol; N.G. and Z.G. optically tracked wave propagation kinetics. N.G. analyzed all experimental data. N.G., A.R.S., and R.L. wrote the manuscript and prepared the figures. N.G., Z.G., A.R.S, and R.L. discussed the results and contributed to the final version of the manuscript.

\bigskip
\textbf{Data Availability}\par
The raw data that support the findings of this study are openly available from the ETH Zurich Research Collection at \url{https://doi.org/10.3929/ethz-c-000787789}.

\bigskip
\textbf{Conflict of Interest}\par
The authors declare no conflict of interest.

\section*{Experimental Section}
\acresetall
\subsection*{Materials}
All chemicals were used as delivered and stored as indicated in the safety data sheets. Potassium \acl{Fi} \acused{Fi} (\ac{Fi}, \qty{98}{\percent}) was purchased from Acros Organics. Ethylenediaminetetraacetic acid (EDTA, \qty{0.5}{\molar} solution at pH 8) and N,N'-methylenebisacrylamide (MBA, 2~w/v\% solution) were purchased from Fisher Bioreagents. 2-Hydroxy-2-methylpropiophenone (HMPP \qty{97}{\percent}) was purchased from Aldrich. Acrylamide (AM, \(\geq\) \qty{99}{\percent}) was purchased from Sigma. D-(+)-Glucose (glucose, \(\geq\) \qty{99.5}{\percent}), alginic acid sodium salt (alginate), \acl{GOx} \acused{GOx}(\ac{GOx} Aspergillus niger, Type VII, batches used \qty{248878}{\enzUnit\per\gram}), sodium hydroxide (NaOH pellets, \(\geq\) \qty{98.0}{\percent}) and mineral oil (light) were purchased from Sigma-Aldrich. Calcium chloride dihydrate (\ce{CaCl2}, \(\geq\) \qtyrange{99.0}{103.0}{\percent}) and hydrochloric acid (HCl, \qty{37}{\percent}) were purchased from VWR Chemicals. Deionized water was used throughout this study unless otherwise mentioned.\par
Analysis was performed using the commercial software MATLAB2023B, unless otherwise stated.

\subsection*{Statistical Analysis}
To assess significant variance between datasets, a bootstrapping methodology was employed. This distribution-independent approach facilitates statistical inference from limited sample sizes.\cite{efron1994IntroductionBootstrap} Significance thresholds were calibrated based on the sample size ($n$) of each dataset. For cases where $n < 6$, a significance level of \qty{75}{\percent} (\qty{25}{\percent}) was established, representing the maximum attainable confidence interval for the median in such instances.\cite{schmid20143vFallacy} It must be noted that all comparisons made in the text would have passed a significance level of \qty{95}{\percent} as well. Conversely, for datasets where $n \ge 6$, a significance level of \qty{95}{\percent} (\qty{5}{\percent}) was utilized, as the increased sample size permits greater statistical certainty.\cite{schmid20143vFallacy} Not significant differences are denoted as n.s.

\subsection*{Stock solutions}
Three stock solutions were prepared for all experiments. First, a \ce{Ca[EDTA]^2-} (CaEDTA) stock solution was prepared by adding a stoichiometric amount of \ce{CaCl2} to an EDTA solution (\qty{0.5}{\molar}). The pH was then adjusted to \num{10 \pm 0.5} with NaOH to prevent immediate crosslinking of alginate upon mixing. Second, an alginate stock solution was prepared by dissolving alginic acid sodium salt powder in water (\qty{45}{\mg} per \qty{1}{\mL}) while stirring and stored at \qty{4}{\celsius} after complete dissolution. Third, an AM stock solution in water was prepared (\qty{50}{\wtPercent} AM) and stored at room temperature.

\subsection*{Rheology}
\label{method:rheopH}
Rheological characterization with in situ pH measurement was conducted to determine the influence of pH on the enzymatically driven gelation of alginate alone (\autoref{fig:mechRheo}). In summary, the rheometer (MCR 302, AntonPaar) was equipped with the standard upper part of a Couette geometry (CC27 Cat. No. 3912, AntonPaar). The bottom cap of the measuring cup was replaced with a custom-made part containing a hole to accommodate a flat-end pH electrode (662-1763, VWR). A custom-made base plate was used to hold the measuring cup. The temperature was not actively controlled as this modification prevented the use of the standard Peltier temperature control unit. Since the rheometer requires a Peltier device to function, it was still connected to the equipment but set aside. Measurements were controlled using the instrument's software (RheoCompass, AntonPaar) in oscillatory mode with a frequency of \qty{1}{\hertz} and a strain of \qty{2}{\percent} for at least \qty{3}{\hour}. Rheological and pH data points were collected independently every \qty{30}{\second} and manually synchronized during data processing. The pH data were acquired using a pH meter (pH 1100 L, VWR) and transmitted to a spreadsheet using a dedicated software plug-in (MultiLab Importer Software, Xylem Analytics). For each measurement, two stock solutions with pH adjusted to \num{10 \pm 0.1} using NaOH were prepared in excess and mixed immediately before measurement in the rheometer's measuring cup at the ratio required to achieve the final concentrations (total volume of \qty{8.4}{\mL}). All concentrations reported are for the final mixed solutions. The first solution contained varying concentrations of CaEDTA, \ac{GOx} (\qty{100}{\enzUnit\per\mL}), and alginate (\qty{2}{\wtPercent}). The second solution contained glucose (\qty{100}{\milli\molar}) and varying concentrations of \ac{Fi}. The average data presented in \autoref{fig:mechRheo}b were calculated from at least two measurements. In \autoref{fig:mechRheo}c, the storage modulus and pH values of equivalent samples were averaged from the last \qty{20}{\minute} of the measurements.

\subsection*{Preparation of Double Network Hydrogels}
\label{method:prepHygel}
To investigate homogeneous enzymatically driven self-stiffening (\autoref{fig:mechTens}), propagation of chemical waves (Figures \ref{fig:waveProp1}, \ref{fig:waveProp2}), and chemomechanical transduction (\autoref{fig:mechTrack}), we prepared two types of \ac{DN} hydrogels, namely adaptive and non-adaptive. The adaptive \ac{DN} hydrogels comprised a permanent \ac{PAM}-based covalent network, a dynamic and pH-responsive CaEDTA-alginate network, and the enzymatic reaction components (\ac{GOx}, \ac{Fi}, and glucose) to drive spatiotemporal pH transitions. Non-adaptive \ac{DN} hydrogels were formulated identically but without the enzymatic mediator (\ac{Fi}) and substrate (glucose), unless otherwise stated. Throughout this study, we fixed the \ac{PAM}:CaEDTA-alginate weight ratio at 85:15. The preparation of \ac{DN} hydrogels followed an adapted protocol based on previously reported \ac{DN} hydrogel systems.\cite{hirsch20213DPrinting, sun2012HighlyStretchable} In general, the \ac{PAM} network was polymerized in the presence of CaEDTA, alginate, and \ac{GOx}, followed by pH adjustment and incorporation of glucose and \ac{Fi} through bath soaking (\autoref{fig:mechTens}d). The polymerization molds were 3D printed from polylactic acid (PLA, various sources) using a commercial 3D printer (i3 MK3S, Prusa). To ensure easy demolding, the mold surfaces were sprayed with a release agent (Smooth-On, Universal Mold Release) and then covered with talcum powder (Puder, Penaten Baby). Excess talcum powder was subsequently removed by brushing and blowing with compressed air. For quasi-static tensile tests, the mold was designed as a slab large enough to fit at least six dogbone-shaped samples with a gauge section measuring \qtyproduct[product-units=power]{21 x 2 x 2}{\mm}. For chemical and mechanochemical wave propagation, the mold was designed as strips measuring \qtyproduct[product-units=power]{21 x 2 x 2}{\mm}. In a typical fabrication, the precursor solution was prepared by mixing AM (\qty{12.5}{\wtPercent}), alginate (\qty{2}{\wtPercent}) and the appropriate amount of CaEDTA in an aqueous solution. To crosslink AM, MBA (\qty{0.4}{\wtPercent}, w.r.t. AM content) and HMPP (\qty{3.4}{\wtPercent}, w.r.t. AM content) were added. To improve \ac{GOx} solubility in adaptive \ac{DN} hydrogels, the enzyme was dissolved in the precursor mixture at the specified concentration before the addition of AM. Next, the precursor solution was pipetted into the molds, which were then placed on a water-cooled metal plate inside a UV chamber. The chamber was flushed with \ce{N2} for \qty{5}{\minute}, after which the precursor solution was irradiated with a UV source equipped with \qty{365}{\nano\meter} LED lamp at \qty{15}{\percent} intensity (DELOLUX 20 A4 equipped with a DELOLUX pilot AxT controller, Delo). The measured intensity at \qty{365}{\nano\meter} was \qty{25}{\milli\watt\per\centi\meter\squared} (measured with DELOLUXcontrol, Delo). After polymerization, the hydrogels were demolded and soaked consecutively in two liquid baths for \qty{10}{\minute} each. The first bath contained equimolar concentrations of CaEDTA (relative to the precursor solution) at pH \num{10 \pm 0.1}, while the second contained the same solution supplemented by glucose (\qty{100}{\milli\molar}) and the appropriate \ac{Fi} concentration. Finally, the hydrogels were removed from the second bath and stored in mineral oil. The acidic hydrogels used to trigger the propagation of chemical waves were comprised of \ac{PAM} alone (\qty{12.5}{\wtPercent}) and polymerized under the conditions described above. After polymerization, trigger hydrogels were stored in an HCl solution adjusted to pH~1.

\subsection*{Tensile Tests}
Quasi-static uniaxial tensile testing was performed on adaptive and non-adaptive \ac{DN} hydrogels after being stored in mineral oil overnight to allow for any spontaneous autocatalytic self-stiffening (\autoref{fig:mechTens}). Dogbone-shaped samples were punched from the hydrogel slab (shape ISO 37, Type 4; punch from Wallace Instruments) and re-immersed in mineral oil until tested. Tensile testing was performed on a universal mechanical testing machine (Z2.5, ZwickRoell) at a crosshead speed of \qty{100}{\mm\per\minute}. A \qty{10}{\newton} load cell (Xforce HP, ZwickRoell) was used, and the samples were secured with screw grips (Screw Grips 8033, ZwickRoell). The initial grip-to-grip separation was maintained constant for all measurements, ensuring that the samples were unloaded at the start. Continuous mechanical deformation was applied until the sample fractured. The strain values were corrected by identifying the point at which the stress began to increase sharply. The mechanical properties were calculated for each sample individually and then averaged for each condition tested ($n = 6\,or\,7$). The apparent modulus and work of fracture were calculated from a linear fit between \qtyrange{50}{100}{\percent} strain and from the area under the curve. Error propagation was used to calculate the standard deviation of the relative increase in mechanical properties from the non-adaptive and adaptive \ac{DN} hydrogel samples.

\subsection*{Optical Tracking of Propagating Chemical Waves}
\label{method:optTrack}
Optical tracking was used to monitor the propagation of the chemical wave fronts in triggered adaptive \ac{DN} hydrogel samples (Figures \ref{fig:waveProp1} and \ref{fig:waveProp2}). The samples were placed in a glass-bottomed mineral oil bath and imaged at regular intervals from above using a reflex camera (EOS 2000D, Canon). For each experiment, eight samples with the same composition were prepared: half were triggered by placing the trigger hydrogel at pH~1 on one end for \qtyrange{1}{5}{\minute}, while the other half were used as reference samples and underwent spontaneous autocatalytic enzymatic reaction (reference samples). First, the samples were individually identified in the images using an open-source software (Fiji, ImageJ)\cite{schindelin2012FijiOpensource} by aligning them horizontally and marking them with a rectangle (Figure \ref{fig:SIimgA}a). Next, the time until the onset of spontaneous autocatalysis was determined using the reference samples, followed by tracking the wave front in the triggered samples from the removal of the trigger hydrogel until the onset of autocatalysis in the reference samples. The propagation speed was obtained by linear regression of the wave front position against propagation time ($n \geq 4$). Further technical information about the optical tracking is provided in the Supporting Information (Figure \ref{fig:SIimgA}).

\subsection*{Tracking Mechanical Waves using Indentation}
\label{method:mechTrack}
\textit{Set up:} The mechanical wave was tracked using indentation, which allows the non-destructive measurement of mechanical properties with spatiotemporal resolution (\autoref{fig:mechTrack}).The indentation measurements were performed using a texture analyzer (TA.XTplus, Stable Micro Systems) equipped with a \qty{500}{\gram} load cell (Stable Micro Systems) and a \qty{1.5}{\mm} radius spherical probe (1/8" diameter, TA-8A, Texture technologies). A manually adjustable x-y stage with \qty{10}{\um} resolution was mounted on the texture analyzer to allow precise and repeatable spatially resolved measurements along the longitudinal direction of the samples. At the start of each spatiotemporal set of measurements, the probe height was set to zero in an empty sample container, which was subsequently filled with mineral oil to host the sample. Three reference sample variations were tested: triggered samples without substrate and mediator (diffusive fronts), untriggered samples with substrate and mediator (spontaneous autocatalytic enzymatic reaction), and untriggered samples without substrate and mediator. All samples remained stationary throughout the experiment. Each indentation measurement was initiated with the probe at a height of \qty{2.7}{\mm}, preventing mechanical contact with the sample. Next, the probe was moved \qty{1.5}{\mm} downwards at a speed of \qty{0.5}{\mm\per\second}, before returning to its original position. For each sample, a baseline measurement was taken ($t = \qty{0}{\minute}$) by indenting the hydrogel along its longitudinal axis before a \qty{30}{\second} or \qty{2}{\minute} trigger. After removing the trigger ($t = \qty{0}{\minute}$), the samples were indented along their longitudinal axis at regular intervals. The measurements were grouped into a single dataset for each time interval. Each dataset was acquired during approximately \qty{10}{\minute}, covering up to \qty{10}{\mm} in distance with a spatial resolution of \qty{200}{\um}. Given that the average propagation speed of the mechanical waves is \qty{4}{\um\per\minute}, the wave front can only move \qty{40}{\um} during the acquisition of a single time interval dataset, which is negligible considering the spatial resolution of the indentation experiment.\par
\textit{Quantification:} We analyzed single force-indentation depth curves by fitting the raw data using the Hertzian model for contact mechanics, with the shear modulus $G$ as a fitting parameter. We assumed a Poisson's ratio of 0.5 and a probe elastic modulus significantly higher than that of the sample. A finite-thickness correction factor $f$ was introduced into the fit to account for the relatively low sample thickness compared to the contact radius.\cite{shull2002ContactMechanics} Uncertainty in determining the contact point for oil-immersed samples was addressed by including two additional fitting parameters: the distance at first contact ($d_0$) and the force at first contact ($F_0$). An approximate contact point was identified using a moving standard deviation approach. Specifically, the standard deviation of the applied force was calculated over a sliding window of 9 consecutive data points. The approximate contact point was defined as the point where this standard deviation first exceeded a threshold value of \qty{0.12}{\mN}. This threshold was selected empirically by analyzing its effect on the quality of regression analysis. Low values resulted in the shear modulus being strongly dependent on threshold selection, while high values resulted in a fitting window that no longer matched the Hertzian model ($F\propto d^{3/2}$). To satisfy the Hertzian approximation, which assumes small strains and contact radii, data fitting was restricted to the regime where the contact radius remained below \qty{20}{\percent} of the actual probe radius. The final fitting function with the incorporated finite-thickness correction factor is: $F = \frac{16}{3}G\cdot R^{1/2}\cdot (d-d_0)^{3/2}\cdot f+F_0$, where $G$ is the shear modulus, $R$ is the probe radius, $d$ is the indentation depth, and $F_0$ and $d_0$ are the distance and force offsets at first contact, respectively. The correction factor $f =(1+1.33(a/h)+1.33(a/h)^3)^{-1}$ accounts for substrate effects in thin samples, where $a/h$ is the ratio of contact radius to sample thickness. The contact radius is given by $a = \sqrt{R\cdot d}$ and the effective sample thickness calculated as $h = h_0-d_0$, with $h_0$ being the initial probe height. Finally, the spatially resolved shear moduli were plotted as a function of their corresponding longitudinal position along the hydrogel. For tracking the mechanical waves, the shear modulus profiles at each time point were fitted with sigmoidal functions, with the baseline as the lower fitting limit for the distal plateau, and the wavefront position was defined as the intersection of the distal plateau with a linear fit of the transition region passing through the inflection point. The overall baseline of the shear modulus was calculated as the mean value of all measurements performed on all samples (adaptive and non-adaptive, triggered and untriggered). The propagation speed was determined with the Theil-Sen estimator, as it is robust towards outliers.

\medskip

\printbibliography

\newpage
\section*{Supporting Figures}
\setcounter{figure}{0}
\renewcommand{\thefigure}{S\arabic{figure}}
\begin{figure}[h!]
        \centering
        \includegraphics[width=\linewidth]{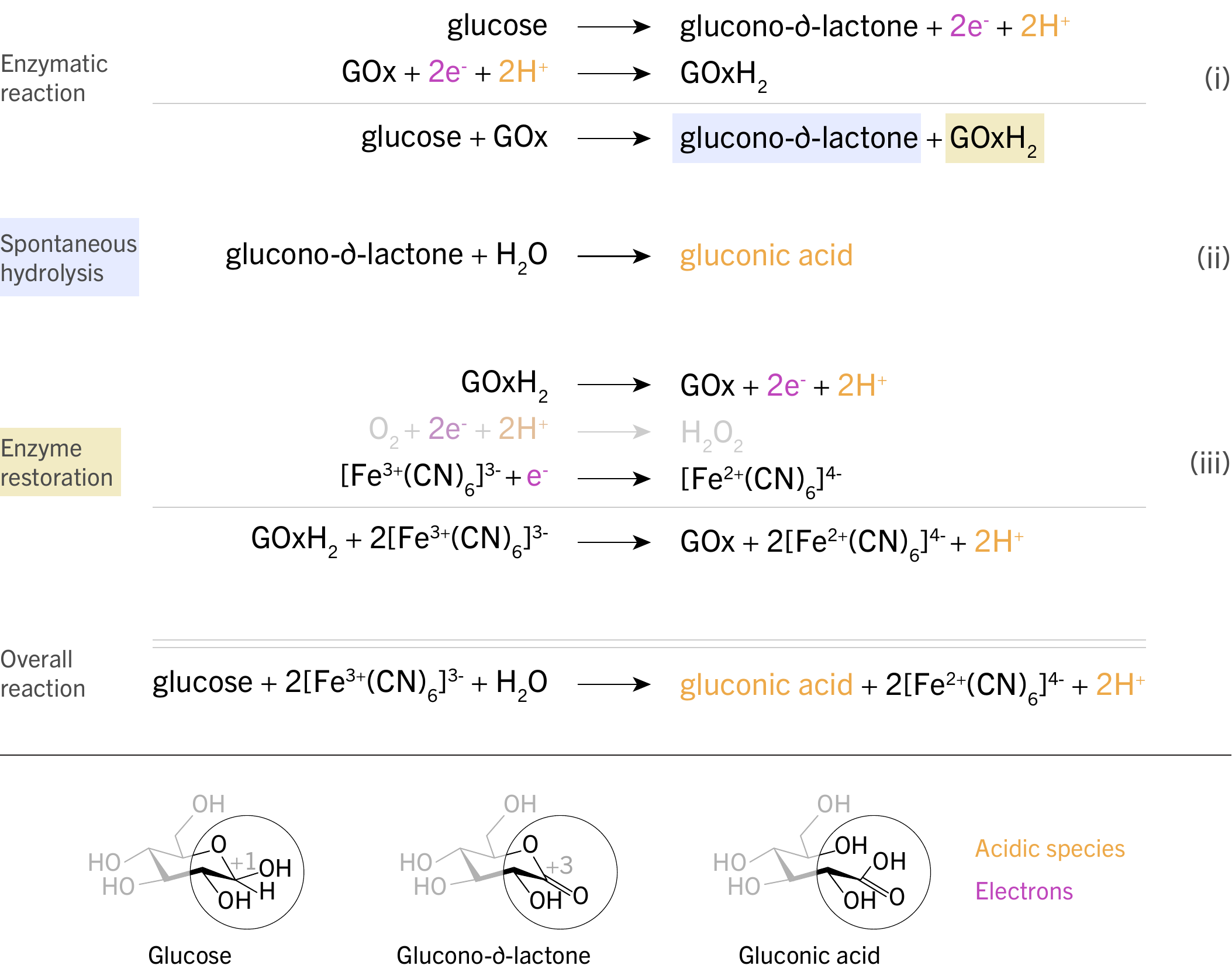}
        \caption{Reaction scheme of the full glucose oxidase (GOx) reaction. (i) Enzymatically-catalyzed oxidation of glucose to glucono-\textdelta-lactone, two electrons and two protons are transferred to the enzyme, which is thus reduced to GOxH\textsubscript{2}. (ii) Glucono-\textdelta-lactone spontaneously hydrolyzes to gluconic acid. (iii) The enzyme is oxidized to its active form by a mediator, typically oxygen, which consumes the protons released in the process. Alternatively, ferricyanide can be reduced to ferrocyanide, which does not react with the protons.}
        \label{fig:SIchemReac}
    \end{figure}

\begin{figure}[h!]
    \centering
    \includegraphics[width=\linewidth]{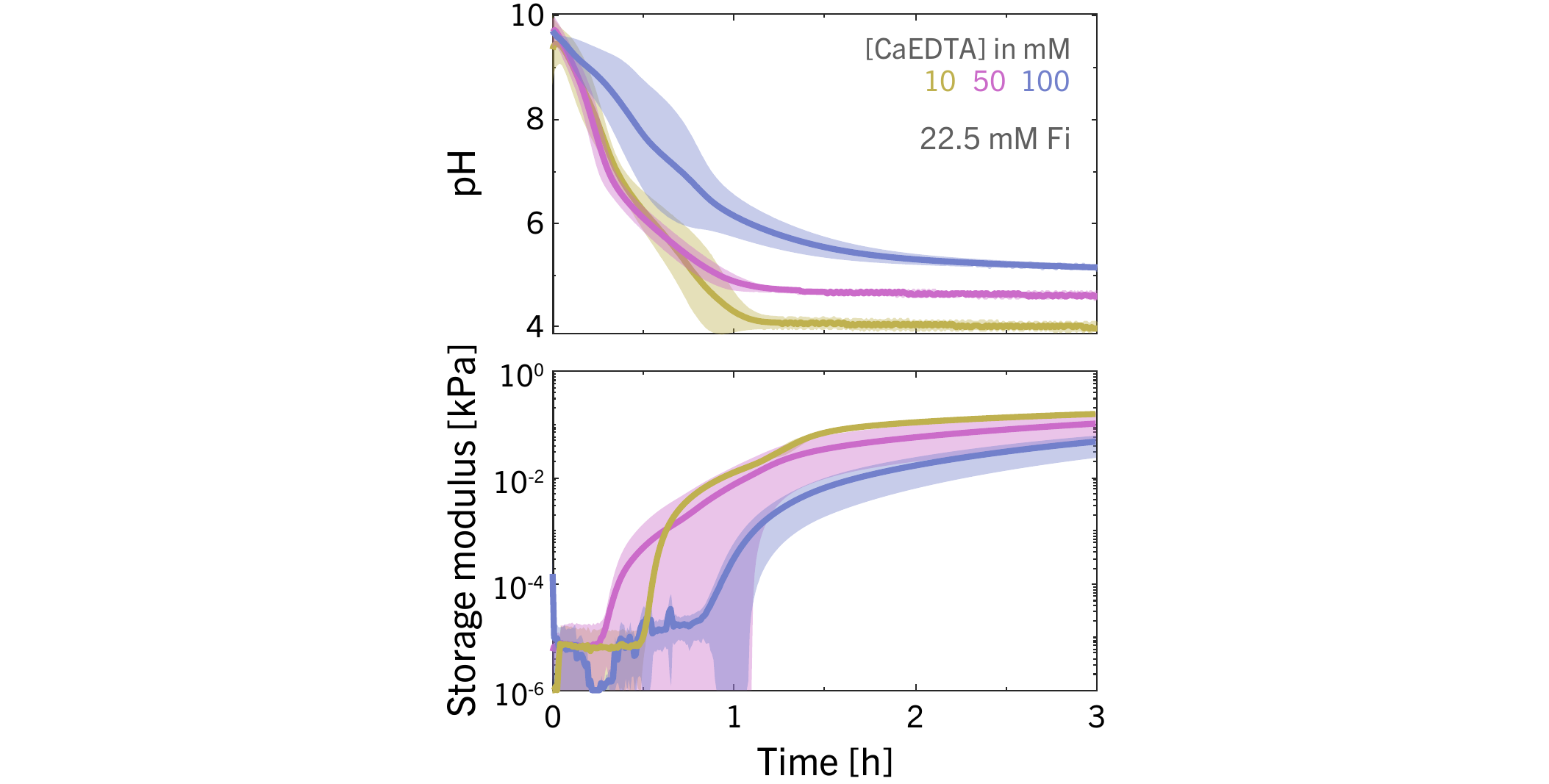}
    \caption{Temporal evolution of pH and storage modulus for \acf{CaEDTA}-alginate hydrogels at three CaEDTA concentrations (\qty{10}{\milli\molar}, \qty{50}{\milli\molar}, \qty{100}{\milli\molar}) with constant \acf{Fi} (\qty{22.5}{\milli\molar}) and glucose (\qty{100}{\milli\molar}). Progressive increase in storage modulus from lower to upper plateau indicates the mechanical transition from uncrosslinked to crosslinked network state. pH measurements and rheological data were collected independently and synchronized during data processing. Solid lines represent averages while shaded regions indicate standard deviation.}
    \label{fig:SIrheoCE}
\end{figure}

\begin{figure}[h!]
    \centering
    \includegraphics[width=\linewidth]{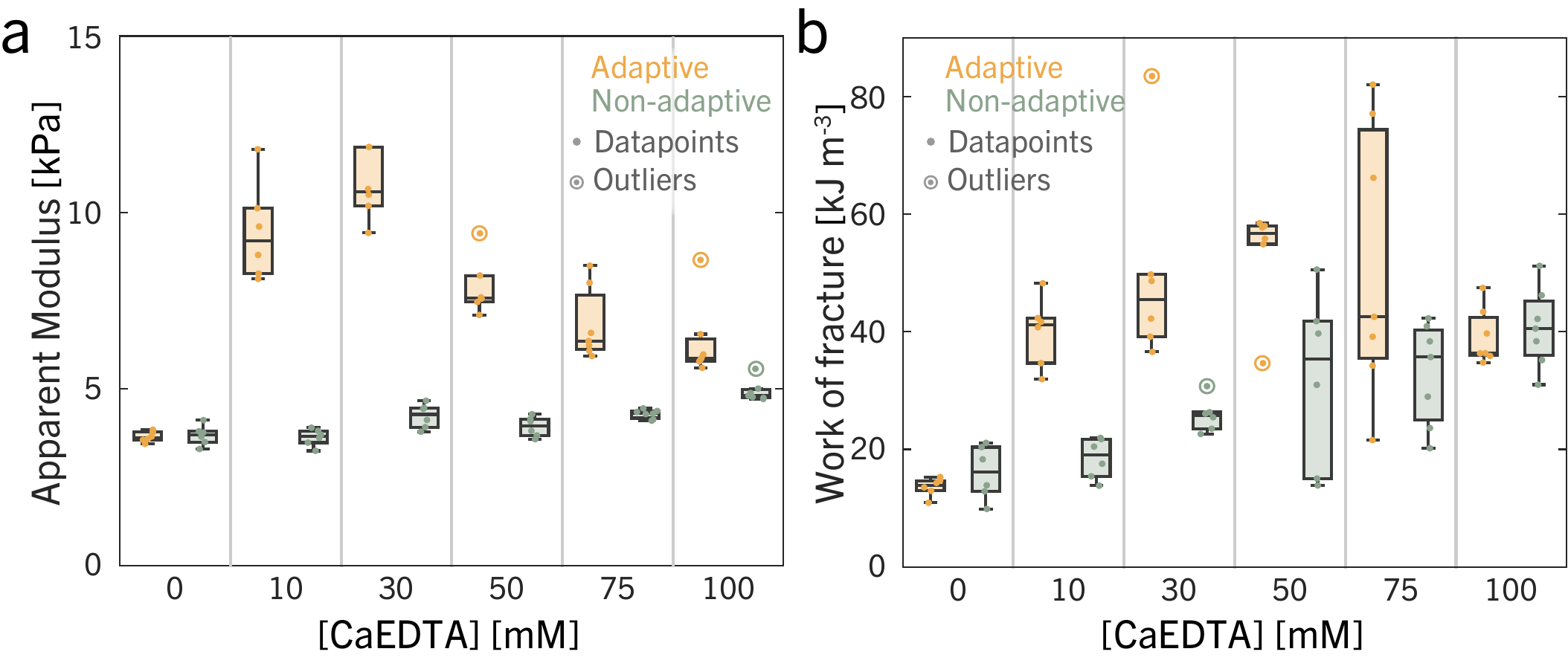}
    \caption{Summary of quasi-static uniaxial tensile testing for self-stiffened adaptive and equilibrated non-adaptive samples. Absolute median and data distribution values of a) apparent modulus and b) work of fracture. Adaptive \acf{DN} hydrogels contained fixed concentrations of \acf{GOx} (\qty{100}{\enzUnit\per\gram}), \acf{Fi} (\qty{22.5}{\milli\molar}), and glucose (\qty{100}{\milli\molar}) whereas non-adaptive \ac{DN} hydrogels contained \ac{GOx} (\qty{100}{\enzUnit\per\gram}). Statistical analysis was done using Matlab2023b}
    \label{fig:SItens}
\end{figure}

\begin{figure}[h!]
    \centering
    \includegraphics[width=\linewidth]{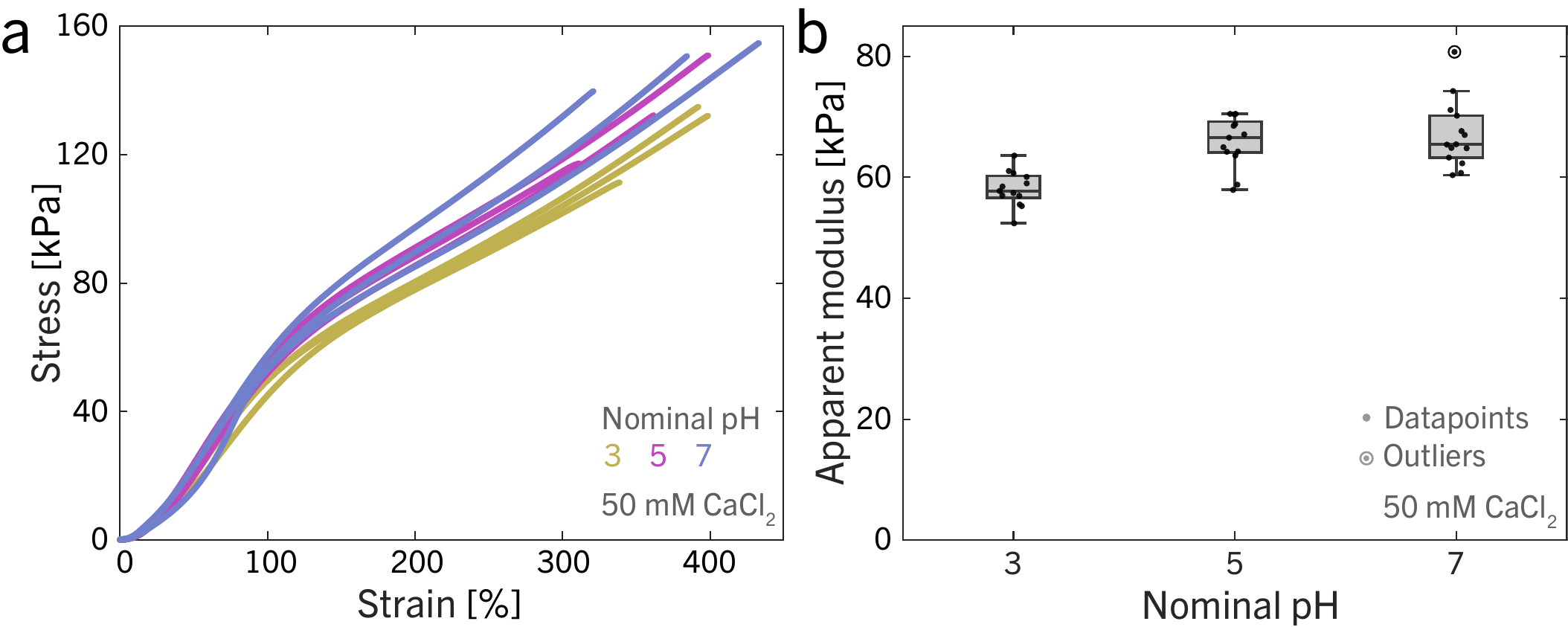}
    \caption{Dependence of apparent modulus in \acl{PAM}-alginate \acf{DN} hydrogels on pH at a constant calcium ion (\ce{Ca^2+}) concentration (\qty{50}{\milli\molar}), determined by quasi-static tensile testing. a) Stress-strain curves of three representative samples for each pH value. b) Resulting apparent modulus at each nominal pH value (experimental values 3.1, 5.2 and 7.3). \ac{DN} hydrogel slabs were prepared as outlined in the Experimental Section of the main text ("Preparation of Double Network Hydrogels") without any enzymatic components and were soaked in baths of \qty{50}{\milli\molar} \ce{CaCl2} at the designated pH value for at least \qty{3}{\hour}. Then, dogbone samples were stamped and re-immersed in the \ce{CaCl2} solution until measured as outlined in the main text ("Tensile tests"). Note that these samples were swollen to equilibrium unlike those in other experiments, limiting the accuracy of direct comparison in absolute values.}
    \label{fig:SItenspH}
\end{figure}

\begin{figure}[h!]
    \centering
    \includegraphics[width=\linewidth]{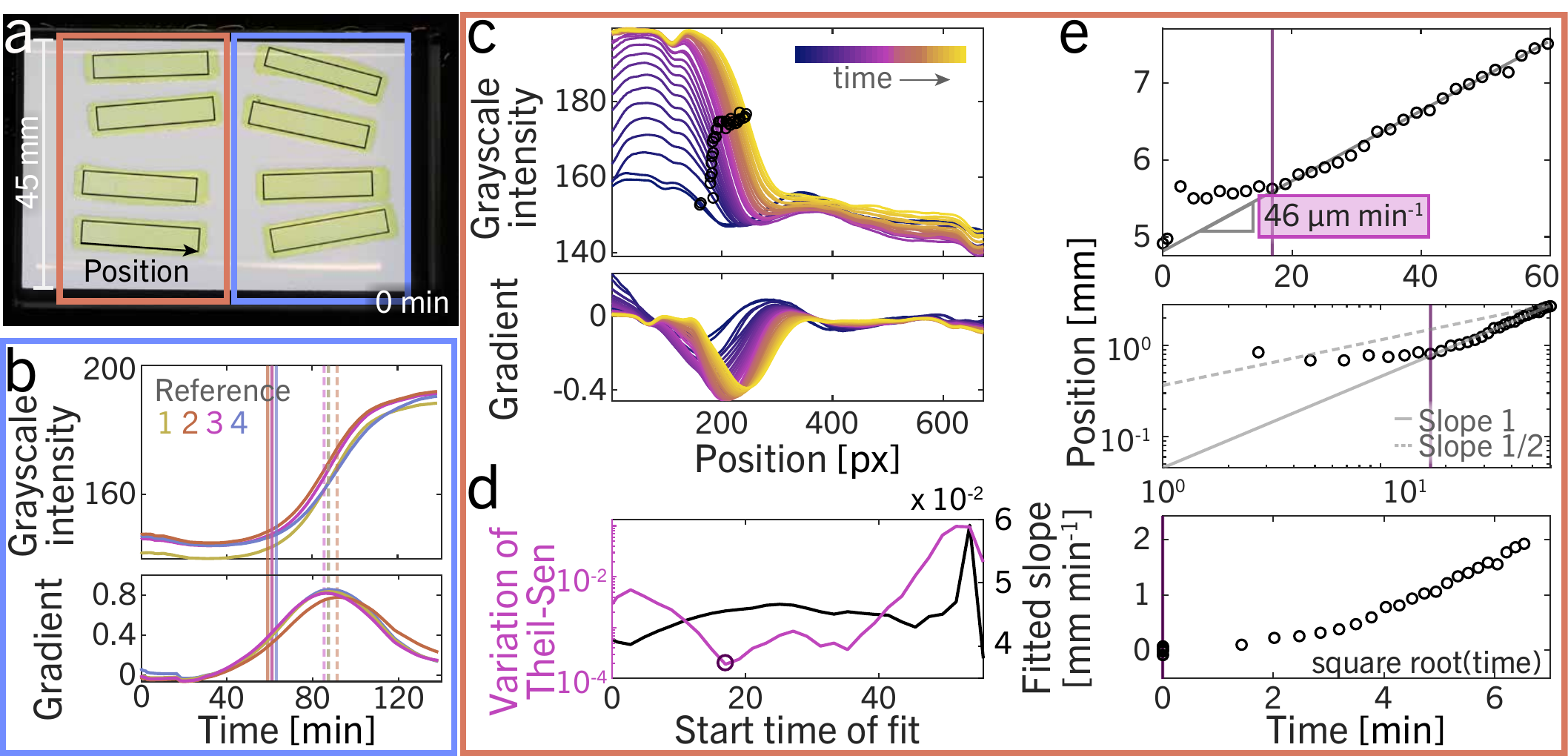}
    \caption{Image analysis methodology for optical tracking of chemical wave propagation. a) Representative image immediately after trigger removal ($t = 0$). Orange rectangle indicate triggered adaptive samples (left side), blue rectangle indicate untriggered adaptive reference samples, and black rectangles define the analysis region for each sample. Sample holder width (\qty{45}{\mm}) provides scale calibration. b) Time-to-autocatalysis calculation from averaged blue-channel intensity of representative samples. Autocatalysis onset is defined as the half-maximum of gradient in grayscale intensity (continuous lines) whereas time-to-autocatalysis is determined by the inflection point of the intensity-time curve, identified as the maximum of the temporal gradient (dashed lines). Image analysis of triggered adaptive samples is interrupted at autocatalysis onset. c) Spatial evolution of blue channel intensity along sample length at sequential time points. Intensity is averaged across the sample width at each longitudinal position. The wavefront position (black circles on top plot) is determined as the inflection point of the spatial intensity profile, corresponding to the minimum of the grayscale intensity gradient (bottom graph). d) Automated detection of linear propagation regime onset. The algorithm iteratively removes data points from the beginning of the dataset and performs linear regression on the remaining data. For each truncated dataset, bootstrapping followed by Theil-Sen regression quantifies fit quality via mean-free relative variance (variance(data-mean)/mean\textsuperscript{2}). The data truncation yielding minimum variance identifies the optimal linear regime and wave propagation onset. Early truncation produces poor fits due to non-linear induction phase and late truncation reduces statistical power. e) Comprehensive data analysis overview. Top: front position versus time with linear fit (continuous line) determining propagation speed (\qty{46}{\um\per\minute}). Middle: log-log plot of front position after y-offset correction from linear fit, demonstrating sensitivity of apparent scaling behavior to data transformation. Bottom: front position versus time, shifted to align with linear regime onset, illustrating deviation from purely diffusive behavior.}
    \label{fig:SIimgA}
\end{figure}

\begin{figure}[h!]
    \centering
    \includegraphics[width=\linewidth]{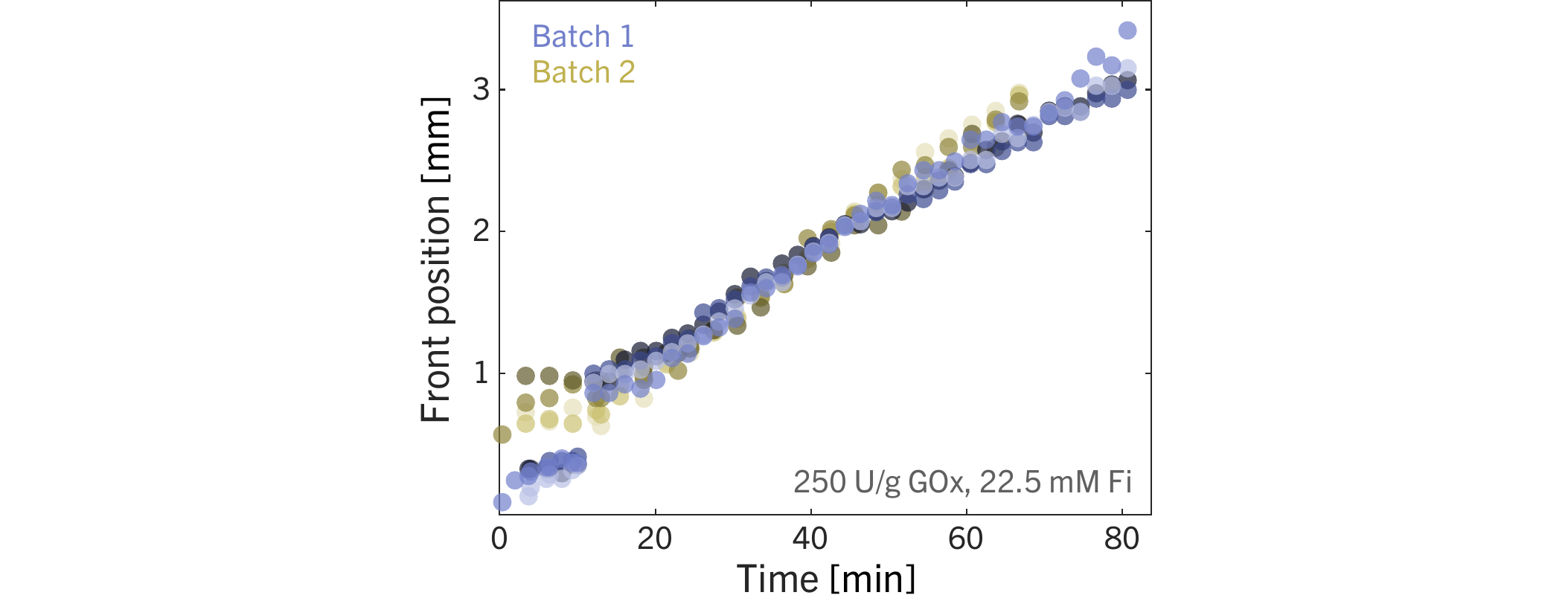}
    \caption{Intra- and inter-batch reproducibility of chemical wave propagation in adaptive \ac{DN} hydrogels. Front positions versus time for two independent batches ($n = 4$ per batch) of adaptive \ac{DN} hydrogels containing  \ac{GOx} (\qty{250}{\enzUnit\per\gram}), \ac{Fi} (\qty{22.5}{\milli\molar}), and glucose (\qty{100}{\milli\molar}). Both batches exhibit consistent and comparable linear propagation speed.}
    \label{fig:SIinterbatch}
\end{figure}

\begin{figure}[h!]
    \centering
    \includegraphics[width=\linewidth]{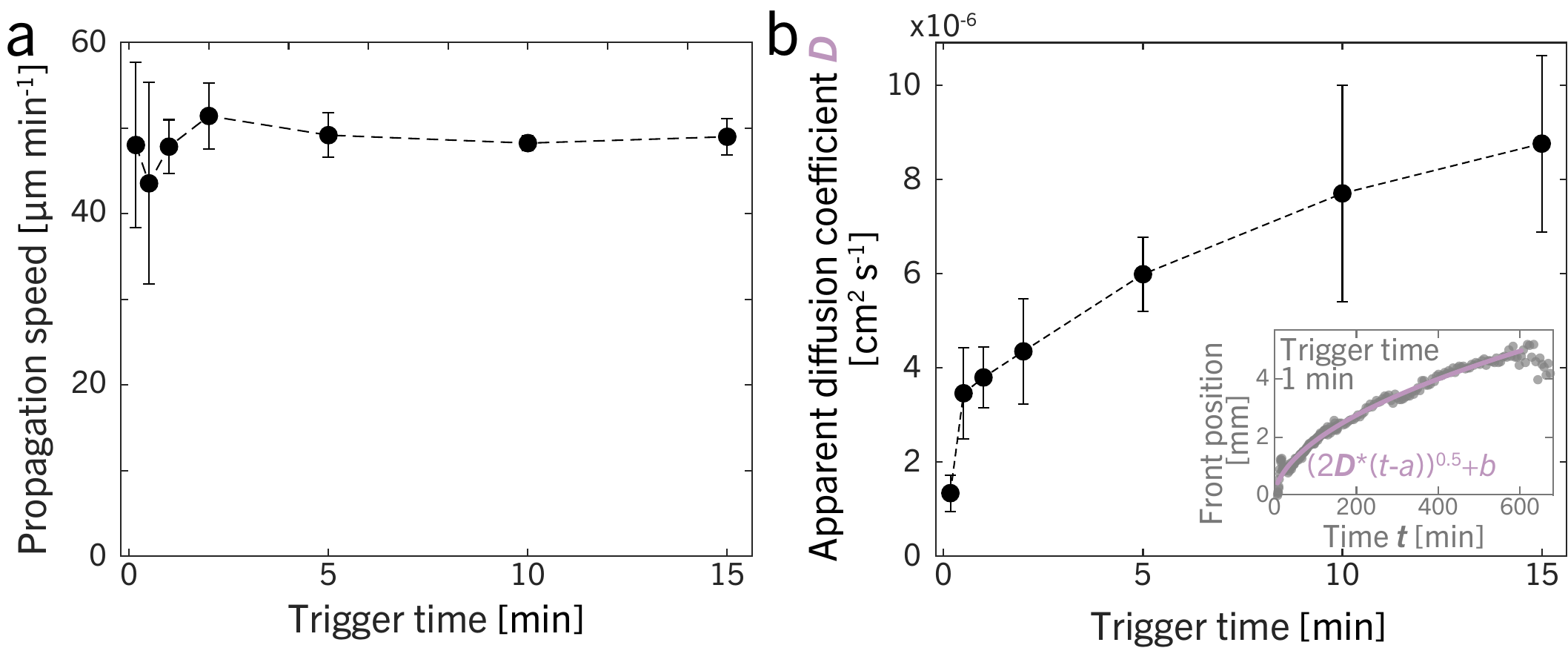}
    \caption{Distinct wave propagation mechanisms in adaptive versus non-adaptive systems as a function of trigger hydrogel contact duration. a) Wave propagation speed in adaptive \ac{DN} hydrogels is independent of trigger time, maintaining  $\approx$ \qty{50}{\um\per\minute}. b) Apparent diffusion coefficient in non-adaptive systems strongly depends on trigger duration, increasing from \qtyrange{1.3\pm 0.4d-6}{8.8\pm 1.9d-6}{\cm\squared\per\second} as trigger time increases from \qtyrange{0.16}{15}{\minute} minutes. Self-sustained front positions were determined from ferricyanide (yellow) conversion into ferrocyanide (colorless) whereas diffusive front positions were determined from Alizarin Red S color change and fitted with diffusive square-root kinetics to obtain the effective diffusion coefficient (inset: $y=\sqrt{2D\cdot(t-a)}-b$, where $a$ and $b$ are fitting parameters). Note that the absolute numbers of $D$ depend on the choice of indicator, these numbers are therefore only accurate in a relative sense.}
    \label{fig:SItrigger}
\end{figure}

\begin{figure}[h!]
    \centering
    \includegraphics[width=\linewidth]{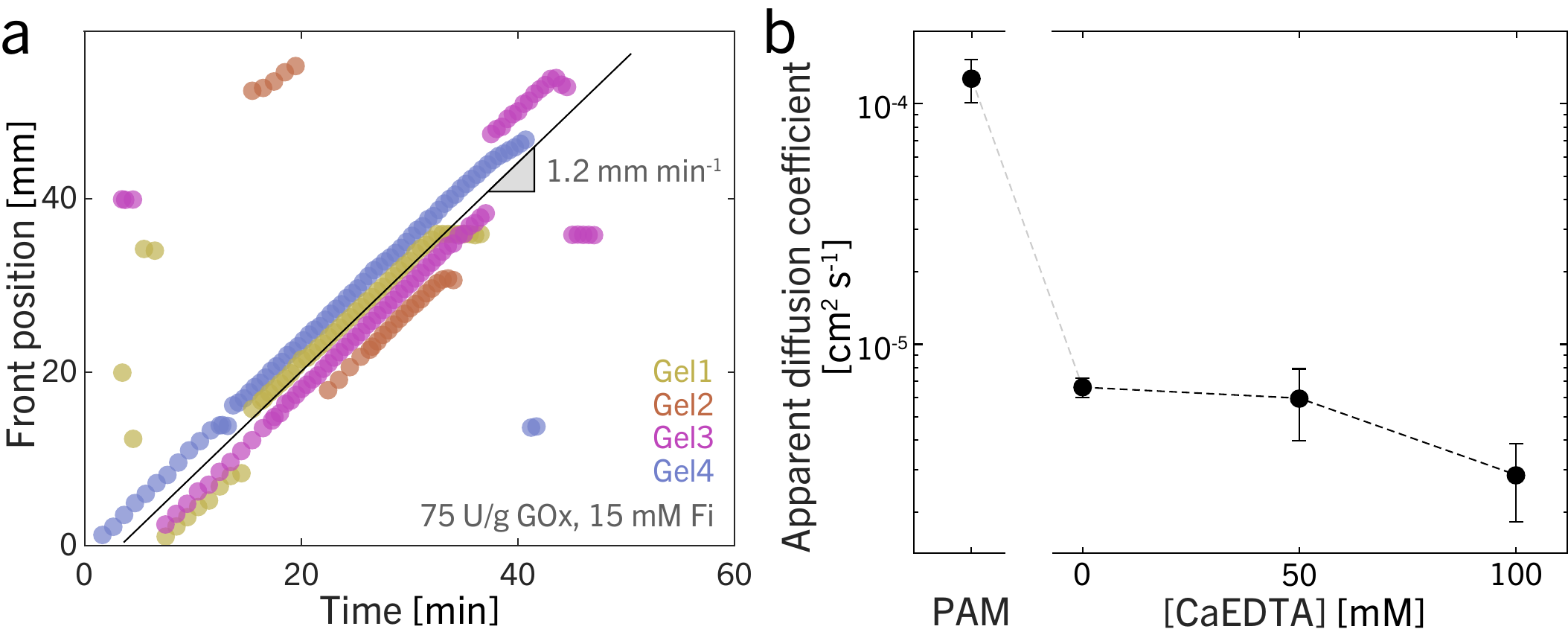}
    \caption{Network complexity and buffering capacity reduces chemical wave propagation speed. a) Wavefront propagation in single-network \acf{PAM} hydrogels (\qty{75}{\enzUnit\per\gram}, \qty{15}{\milli\molar} \ac{Fi}) proceeds at $\sim\qty{1.2}{\mm\per\minute}$, approximately \numrange{25}{70}-fold faster than in \ac{DN} hydrogels (from \qtyrange{15}{44}{\um\per\minute}). Four replicate samples shown with shifted coordinates for clarity. b) Effective \ce{H+} diffusion coefficient in \ac{PAM}-only systems versus \ac{PAM}-alginate \ac{DN} hydrogels with varying CaEDTA concentrations (buffering capacity). \ac{DN} hydrogels exhibit ($D^{H+}_{app}$ by more than one order of magnitude lower than single-network hydrogels due to increased network tortuosity. The system's buffering capacity provides a secondary contribution to the reduction in ($D^{H+}_{app}$ by maintaining \ce{H+} in bound states. Note that the absolute numbers of ($D^{H+}_{app}$ depend on the choice of the pH indicator, enabling only relative comparisons between datasets acquired using the same experimental protocol.}
    \label{fig:SIpam}
\end{figure}

\begin{figure}[h!]
    \centering
    \includegraphics[width=\linewidth]{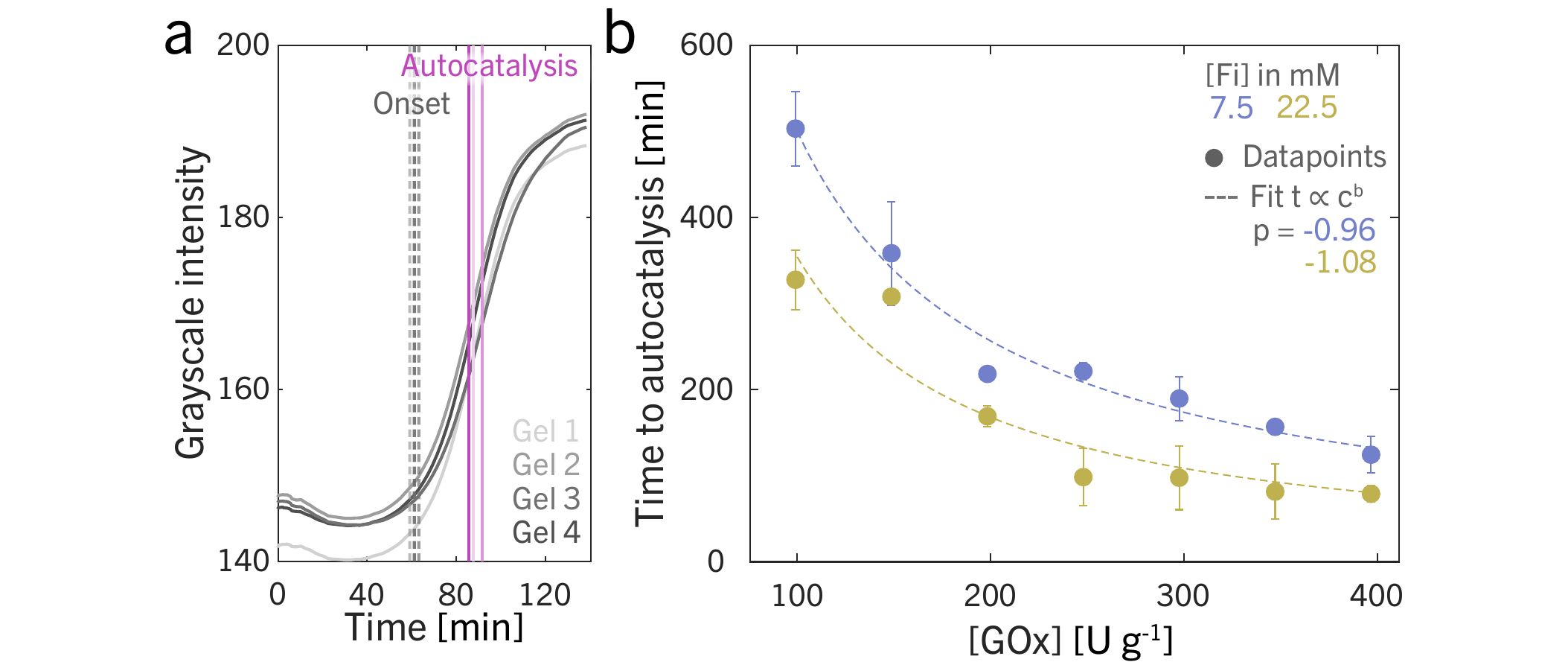}
    \caption{Quantifying spontaneous \ac{GOx} autocatalysis in adaptive \ac{DN} hydrogels. a) Averaged grayscale intensity evolution over time in adaptive \ac{DN} hydrogels ($n = 4$). Time-to-autocatalysis is calculated at the inflection point of intensity curves (continuous magenta lines), where the temporal gradient is at maximum. Autocatalysis onset is defined as the half-maximum of gradient in grayscale intensity (dashed black lines).b) Power law dependence ($y=a\cdot x^p$) of time-to-autocatalysis on \ac{GOx} concentration for $Fi = \qty{7.5}{\milli\molar}$ (blue) and $\qty{7.5}{\milli\molar}$ (yellow). Scaling exponents ($p = -0.96$ and $-1.08$) indicate that doubling \ac{GOx} concentration halves the time-to-autocatalysis. The higher sensitivity compared to wave propagation speed ($p \approx 0.5-0.7$, Figure~5b) explains the inverse correlation between speed and time-to-autocatalysis.}
    \label{fig:SIautocat2}
\end{figure}

\begin{figure}[h!]
    \centering
    \includegraphics[width=\linewidth]{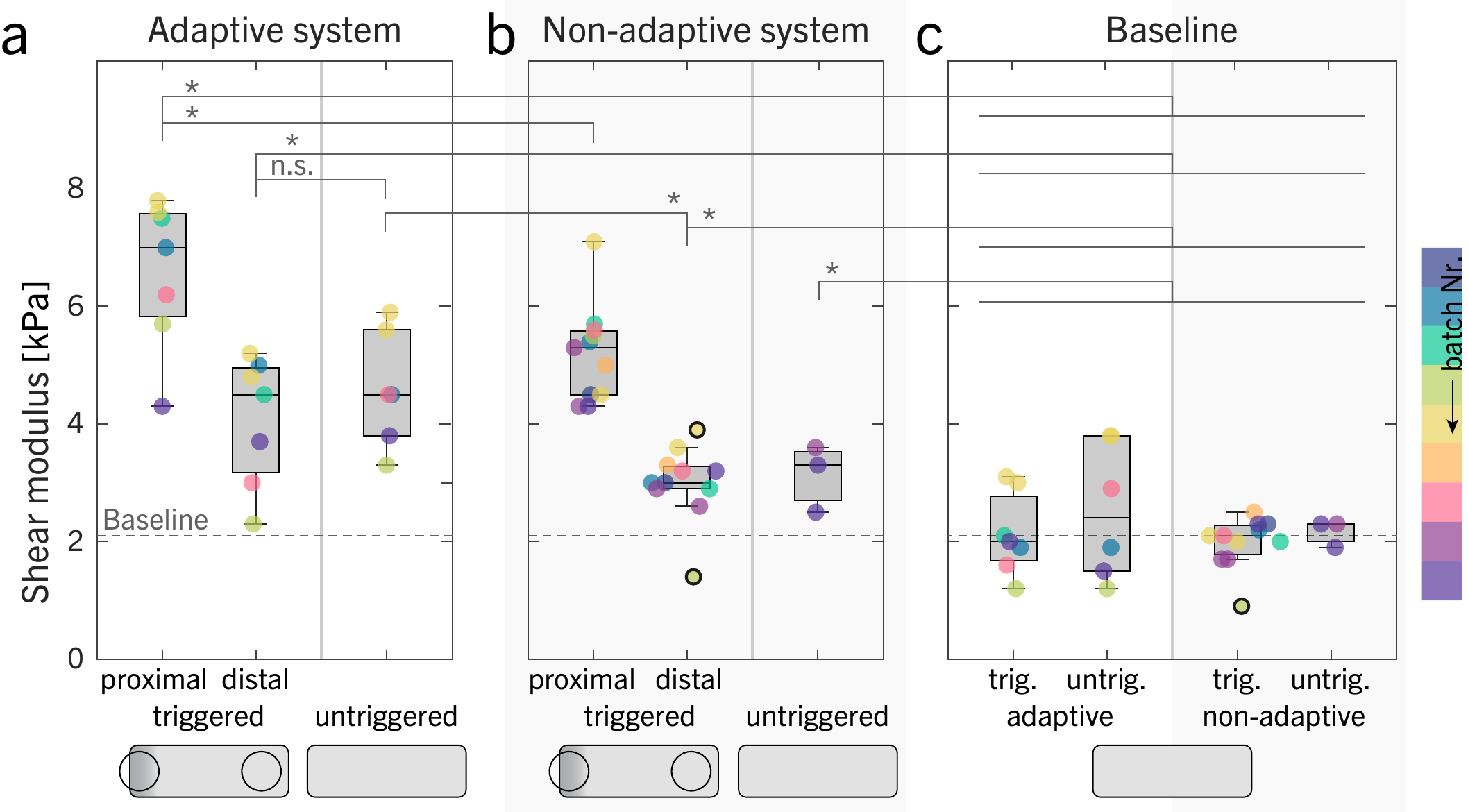}
    \caption{Statistical distribution of shear moduli from spatiotemporally resolved indentation measurements. a) Adaptive \ac{DN} hydrogels exhibit gradient in shear modulus. Triggered regions reach median values of \qty{7.0}{\kilo\pascal} (\acf{IQR} \qtyrange{5.8}{7.6}{\kilo\pascal}), distal regions of triggered samples reach \qty{4.5}{\kilo\pascal} (IQR \qtyrange{3.2}{5.0}{\kilo\pascal}), and untriggered reference samples have an overall shear modulus of \qty{4.5}{\kilo\pascal} (IQR \qtyrange{3.8}{5.6}{\kilo\pascal}). b) Non-adaptive \ac{DN} hydrogels show no enzymatic stiffening with triggered regions at \qty{5.3}{\kilo\pascal} (IQR \qtyrange{4.5}{5.6}{\kilo\pascal}) and distal regions at \qty{3.0}{\kilo\pascal} (IQR \qtyrange{2.9}{3.3}{\kilo\pascal}), untriggered reference samples have an overall shear modulus of \qty{3.3}{\kilo\pascal} (IQR \qtyrange{2.7}{3.5}{\kilo\pascal}). c) Summary of pre-triggered baseline shear moduli for all experimental conditions tested, the overall baseline has a shear modulus of \qty{2.1}{\kilo\pascal}, IQR \qtyrange[range-phrase=~--~]{1.8}{2.3}{\kilo\pascal}. Box plots represent batch distributions (color-coded) with boxes spanning interquartile range. Batches 1-4: \qty{50}{\milli\molar} CaEDTA, \qty{5}{\minute} trigger duration. Batches 5-9: \qty{30}{\milli\molar} CaEDTA and were triggered for \qtyrange{0.5}{2}{\minute}. * denote statistical significance determined as outlined in the Experimental Section for $n \ge 6$, only comparisons mentioned in the text are shown.}
    \label{fig:SImTG}
\end{figure}

\begin{figure}[h!]
    \centering
    \includegraphics[width=\linewidth]{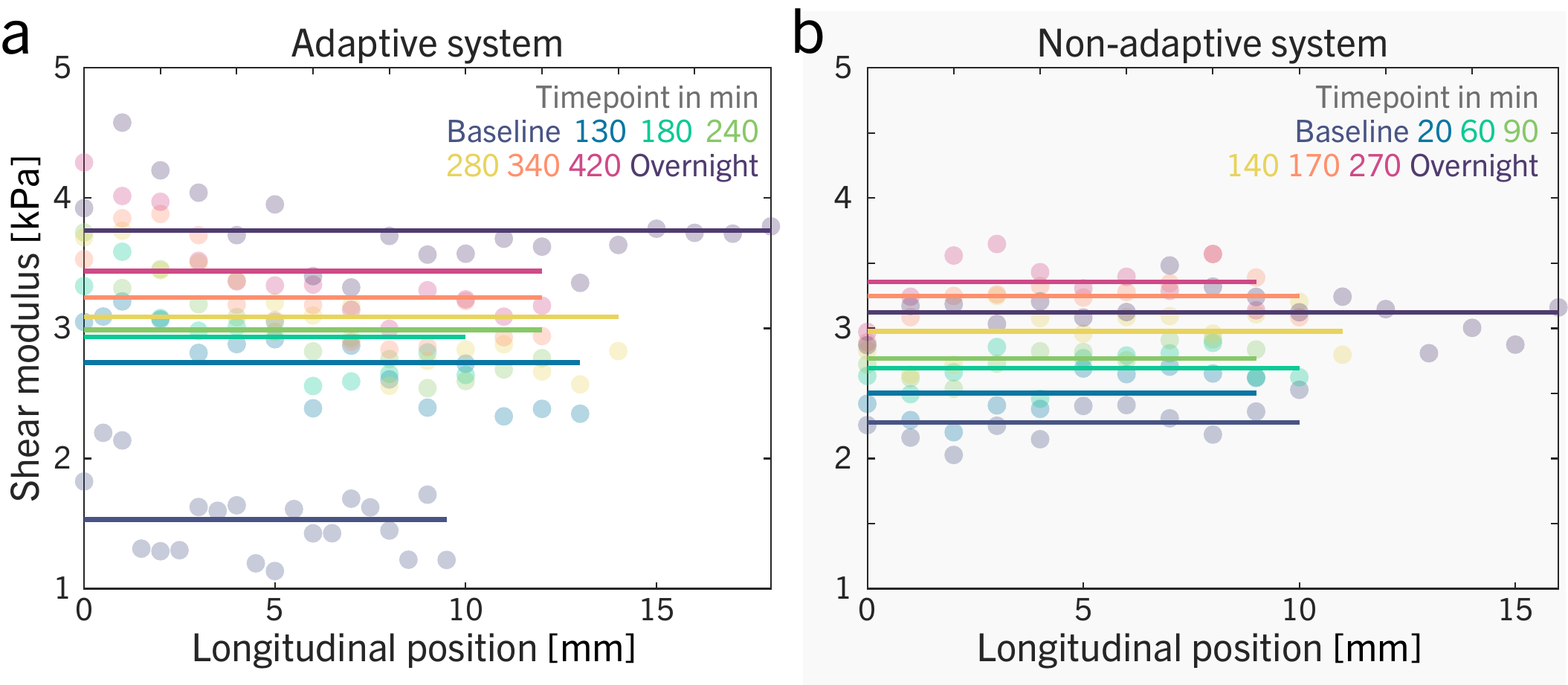}
    \caption{Spatiotemporal shear modulus evolution in untriggered reference samples. a) Adaptive \ac{DN} hydrogels undergo homogeneous temporal stiffening via spontaneous \ac{GOx} autocatalysis activation, with shear modulus increasing from baseline ($\sim \qty{1.5}{\kilo\pascal}$) to quasi-steady-state values ($\sim\qty{3.75}{\kilo\pascal}$ at overnight timepoints) uniformly across the samples' longitudinal axis. The absence of spatial gradients contrasts with triggered samples (Figure~6), confirming that localized chemical signaling generated gradients in mechanical properties. b) Non-adaptive \ac{DN} hydrogels exhibit minimized temporal variation, validating that enzymatic activity is the major driving force for the self-stiffening in these adaptive \ac{DN} hydrogels. Horizontal lines represent average shear modulus calculated from all spatial measurements at each timepoint. Individual datapoints indicate position-resolved measurements.}
    \label{fig:SImTref}
\end{figure}

\begin{figure}[h!]
    \centering
    \includegraphics[width=\linewidth]{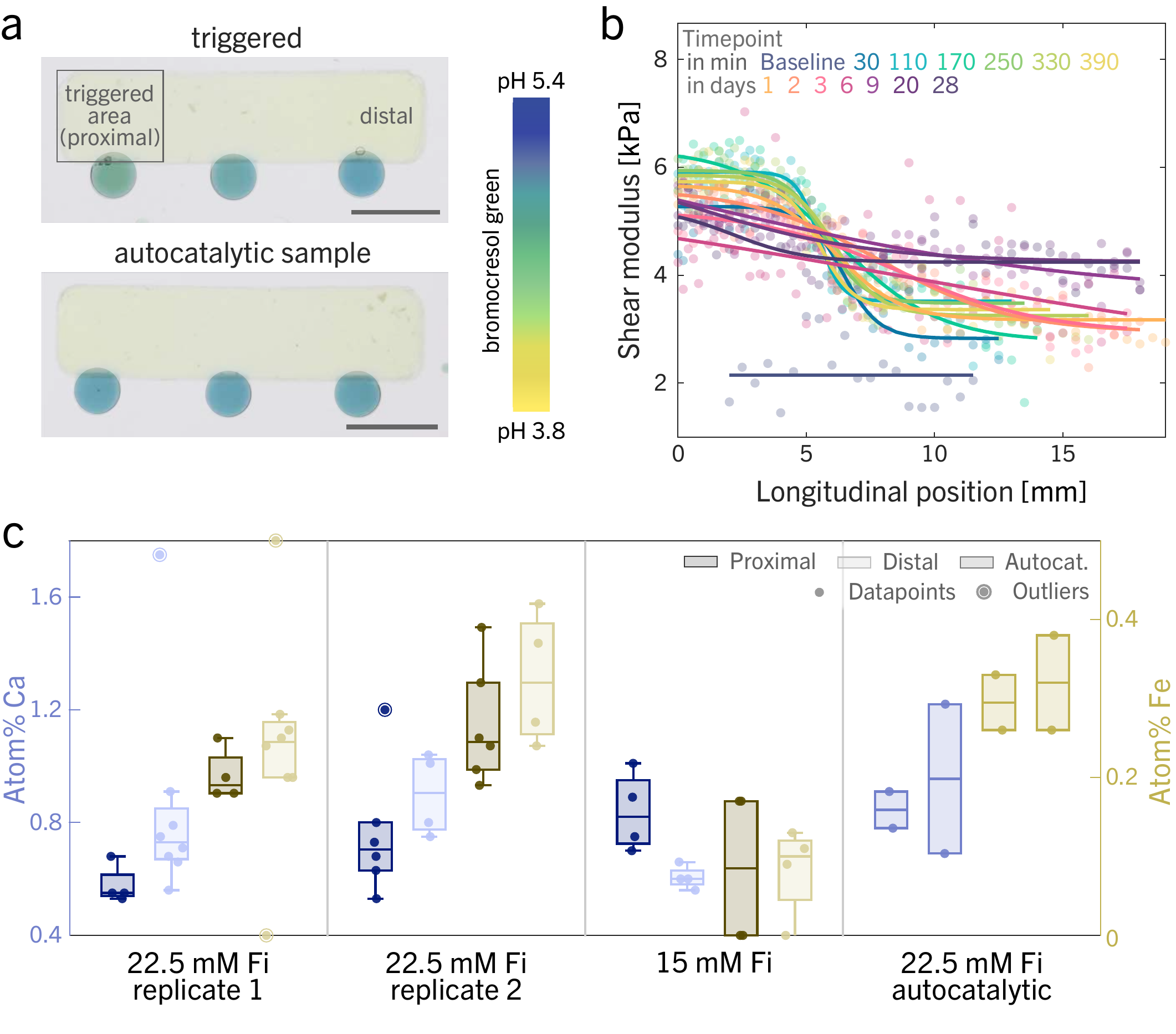}
    \caption{Correlation between persistent pH, mechanical gradients, and calcium concentration. a)~Bromocresol green pH indicator (color range pH 3.8-5.4, shown at full saturation) applied to adaptive \ac{DN} hydrogels 1~day post-triggering. The triggered sample (top) exhibits a spatial pH gradient (pH $\sim 4.6$ at triggered region and pH $\sim 5.0$ at distal region), whereas the spontaneously stiffened reference sample (bottom) shows uniform pH $\sim 5.0$. The observed pH range (4.6-5.0) corresponds to the steep region of the pH-dependent crosslinking curve (Figure~2c) where mechanical properties vary most sensitively with pH. Scale bars: \qty{5}{\mm}. b) Longitudinal shear modulus profiles measured over 28~days in a non-adaptive \ac{DN} hydrogel following acidic trigger removal. The mechanical gradient relaxes gradually due to slow \ce{H+} diffusion, remaining detectable even after 6~days, after which it comparable to untriggered samples as in Figure~\ref{fig:SImTref}. Similar behavior is observed in adaptive samples at long time scales. Colored lines indicate sequential timepoints. Scattered points show individual spatiotemporally-resolved measurements. c)~Calcium (Ca) and iron (Fe) concentrations at either end of four different samples determined by energy-dispersive X-ray spectroscopy (EDX). The proximal and distal ends of triggered sample were analyzed independently, with data points acquired at multiple locations. If significant \ce{Ca^2+} diffusion and chelation with alginate carboxylate groups occurred during chemical wave propagation, a Ca concentration gradient would be expected between proximal and distal regions. However, the EDX measurements reveal no statistically significant Ca concentration difference between these regions (tested as per the Experimental Section), suggesting that such accumulation does not occur to a detectable extent. Additionally, these measurements demonstrate detection sensitivity for Fi concentration differences as low as \qty{15}{\milli\molar}. It should be noted that EDX detection limits are strongly dependent on atomic number, therefore, the sensitivity observed for Fe (Z=26) cannot be directly extrapolated to Ca (Z=20). Adaptive samples were cut in half to separate the the proximal and distal ends approximately \qty{20}{\hour} after triggering the enzymatic reaction, frozen in liquid nitrogen and freeze dried (Labconco Freezone 2.5). Subsequently, the samples were coated with a \qty{6}{\nm} platinum layer and analyzed with EDX in a Scanning Electron Microscope (Gemini 450, Zeiss).}
    \label{fig:SIpHLongEDX}
\end{figure}

\begin{figure}[h!]
    \centering
    \includegraphics[width=\linewidth]{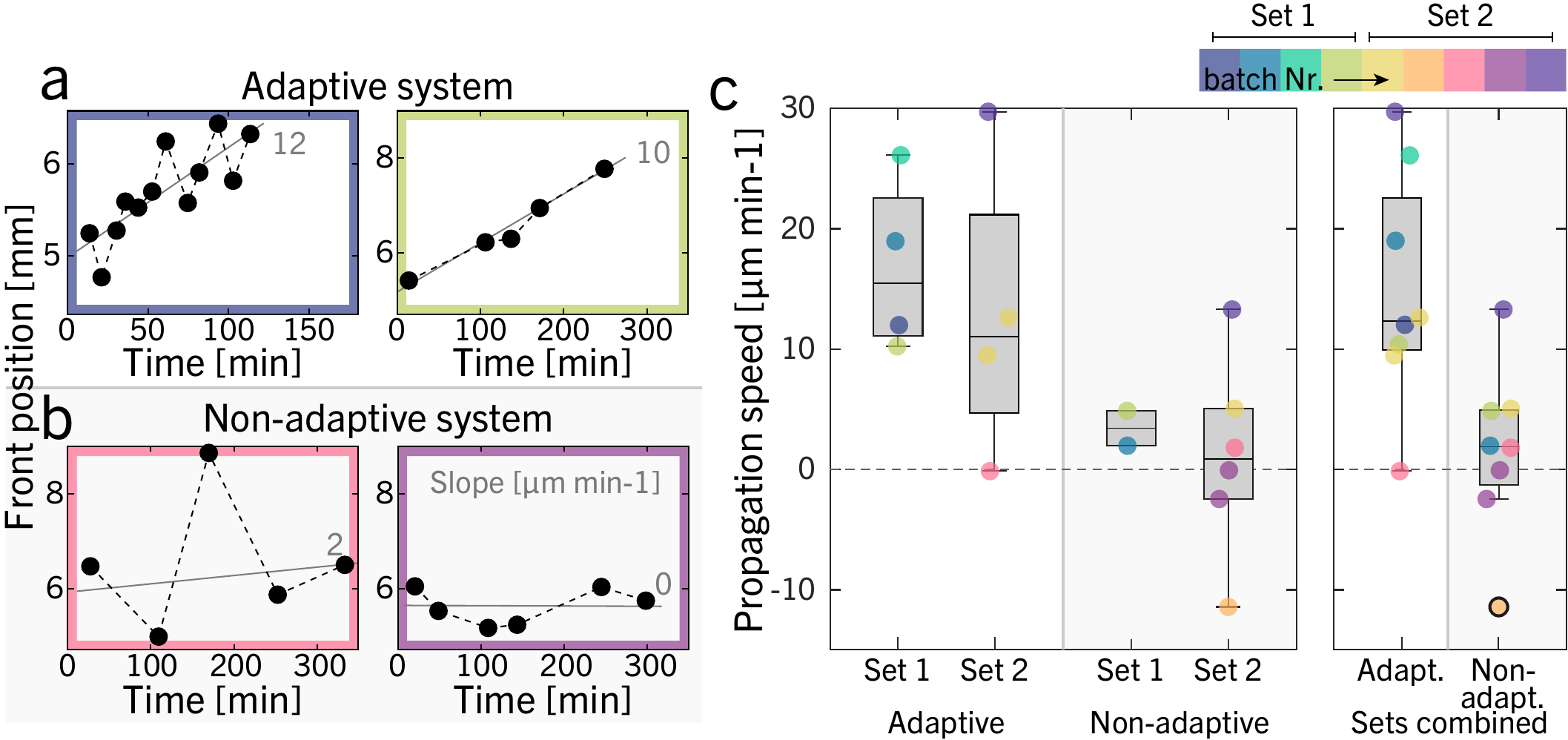}
    \caption{Mechanical wave propagation speeds. a-b) Temporal evolution of the mechanical wavefront position in adaptive (a) and non-adaptive (b) \acf{DN} hydrogels. The mechanical wave propagation speed is determined from the pre-autocatalysis phase prior to \qty{300}{\minute} shown here, where self-sustained chemical waves were observed to propagate with constant speed. Propagation speeds were determined from Theil-Sen linear fits to wavefront position (onset of the distal plateau) versus time for all triggered samples. To facilitate direct comparison, the first panel in (a) maintains an identical axis scaling to the other panels. c) Robustness of mechanical wave propagation across experimental parameters and different batches. Left: Comparison between Set 1 (batches 1-4: \qty{50}{\milli\molar} CaEDTA, \qty{5}{\minute} trigger) and Set 2 (batches 5-9: \qty{30}{\milli\molar}, \qtyrange{0.5}{2}{\minute} trigger). Adaptive \ac{DN} hydrogels exhibit consistent median propagation speed across both parameter sets, confirming that mechanical wave propagation is robust to small variations in CaEDTA concentration and trigger duration. Non-adaptive systems show small positive (Set 1) or near-zero slopes (Set 2) reflecting limited diffusive propagation. Right: Combined data demonstrate clear distinction dynamic behavior between adaptive (\qty{12}{\um\per\minute} IQR \qtyrange{10}{23}{\um\per\minute}) and non-adaptive (\qty{2}{\um\per\minute} IQR \qtyrange{-1}{4}{\um\per\minute}) systems. Colors indicate experimental batches. Data in the main text stems from batch five (yellow) for the adaptive system and batch eight (light purple) for the non-adaptive system.}
    \label{fig:SImTSpeedsFull}
\end{figure}

\end{document}